\documentclass[12pt]{article}
\usepackage[latin1]{inputenc}

\usepackage{amsmath}
\usepackage{color}
\usepackage{amsfonts}
\usepackage{amssymb}
\usepackage[mathscr]{euscript}
\usepackage{graphicx}
\graphicspath{ {./Figures/} } 
\usepackage{geometry}
\usepackage{epsfig}
\usepackage{hyperref}
\usepackage{comment}
\usepackage{tabularx}
\usepackage{bm}
\usepackage{euscript}
\usepackage{graphicx}
\usepackage{exscale}
\usepackage{amsbsy}
\usepackage{subfigure}
\usepackage{textcomp}
\usepackage{comment}
\usepackage{slashed}
\usepackage{authblk}
\usepackage{extarrows}
\usepackage[normalem]{ulem} 
\usepackage{dsfont} 

\usepackage[dvipsnames]{xcolor}
\usepackage{exscale}
\usepackage{amsbsy}
\usepackage{textcomp}
\usepackage{wrapfig}
\usepackage[font=footnotesize,labelfont=bf]{caption}
\usepackage{nicefrac}

\usepackage{braket}
\usepackage{mathrsfs} %
\usepackage{pifont}

\usepackage[outline]{contour}

\def\ba#1\ea{\begin{align}#1\end{align}}
\def\bg#1\eg{\begin{gather}#1\end{gather}}
\def\bm#1\em{\begin{multline}#1\end{multline}}
\def\bmd#1\emd{\begin{multlined}#1\end{multlined}}

\newcommand{\be}{\begin{equation}}
\newcommand{\ee}{\end{equation}}
\newcommand{\bea}{\begin{eqnarray}}
\newcommand{\eea}{\end{eqnarray}}

\newcommand{\pd}{\partial}

\newcommand{\matleft}{\left(\begin{array}}
\newcommand{\matright}{\end{array}\right)}

\newcommand{\sgn}{\operatorname{sgn}}

\usepackage[numbers,sort&compress]{natbib}
\def\simge{
    \mathrel{\rlap{\raise 0.511ex 
        \hbox{$>$}}{\lower 0.511ex \hbox{$\sim$}}}}

\def\simle{
    \mathrel{\rlap{\raise 0.511ex 
        \hbox{$<$}}{\lower 0.511ex \hbox{$\sim$}}}}

\makeatletter
\def\blfootnote{\xdef\@thefnmark{}\@footnotetext}
\makeatother

\hypersetup{
    bookmarks=true,         
    unicode=false,          
    pdftoolbar=true,        
    pdfmenubar=true,        
    pdffitwindow=false,     
    pdfstartview={FitH},    
    pdftitle={Collusion of Interactions and Disorder at the SF-I Transition},    
    pdfauthor={},     
    pdfsubject={Subject},   
    pdfcreator={},   
    pdfproducer={}, 
    pdfkeywords={keyword1} {key2} {key3}, 
    pdfnewwindow=true,      
    colorlinks=true,       
    linkcolor=OliveGreen, 
    citecolor=NavyBlue,        
    filecolor=magenta,      
    urlcolor=cyan           
} 

\makeatletter
\renewcommand\section{\@startsection {section}{1}{\z@}%
                                 {-3.5ex \@plus -1ex \@minus -.2ex}
                                   {2.3ex \@plus.2ex}%
                                   {\normalfont\large\bfseries}}
\renewcommand\subsection{\@startsection{subsection}{2}{\z@}%
                                   {-3.25ex\@plus -1ex \@minus -.2ex}%
                                     {1.5ex \@plus .2ex}%
                                     {\normalfont\bfseries}}
\renewcommand\subsubsection{\@startsection{subsubsection}{3}{\z@}%
                                   {-3.25ex\@plus -1ex \@minus -.2ex}%
                                     {1.5ex \@plus .2ex}%
                                     {\normalfont\itshape}}
\makeatother

\def\pplogo{\vbox{\kern-\headheight\kern -29pt
\halign{##&##\hfil\cr&{\ppnumber}\cr\rule{0pt}{2.5ex}&\ppdate\cr}}}
\makeatletter
\def\ps@firstpage{\ps@empty \def\@oddhead{\hss\pplogo}%
  \let\@evenhead\@oddhead 
}
\def\maketitle{\par
 \begingroup
 \def\thefootnote{\fnsymbol{footnote}}
 \def\@makefnmark{\hbox{$^{\@thefnmark}$\hss}}
 \if@twocolumn
 \twocolumn[\@maketitle]
 \else \newpage
 \global\@topnum\z@ \@maketitle \fi\thispagestyle{firstpage}\@thanks
 \endgroup
 \setcounter{footnote}{0}
 \let\maketitle\relax
 \let\@maketitle\relax
 \gdef\@thanks{}\gdef\@author{}\gdef\@title{}\let\thanks\relax}
\makeatother

\numberwithin{equation}{section}

\DeclareMathAlphabet{\mathcald}{U}{dutchcal}{m}{n}
\SetMathAlphabet{\mathcald}{bold}{U}{dutchcal}{b}{n}

\let\osl\o
\renewcommand{\o}{\over}

\newcommand{\nt}{\notag\\}
\newcommand{\eq}[1]{\begin{align}#1\end{align}}
\newcommand{\abs}[1]{\left| #1 \right|}
\renewcommand{\(}{\left(}
\renewcommand{\)}{\right)}
\renewcommand{\[}{\left[}
\renewcommand{\]}{\right]}
\let\ptl\partial
\renewcommand{\dag}{\dagger}

\renewcommand{\v}{\boldsymbol}
\newcommand{\vb}{\mathbf}
\newcommand{\vx}{{\mathbf{x}}}
\newcommand{\vq}{{\mathbf{q}}}
\newcommand{\vp}{{\mathbf{p}}}

\newcommand{\ep}{\epsilon}

\renewcommand{\b}{\beta}
\renewcommand{\d}{\delta}
\renewcommand{\t}{\tau}
\newcommand{\s}{\sigma}

\newcommand{\V}{{\mathcal{V}}}

\renewcommand{\O}{{\mathcal{O}}}

\newcommand{\M}{{\mathcal{M}}}
\newcommand{\N}{\mathcal{N}}

\newcommand{\T}{{\mathcald{T}}}
\newcommand{\PH}{{\mathcald{PH}}}
\newcommand{\bDelta}{{\bar{\Delta}}}

\newcommand{\col}{\color{RedViolet}\bf}
\newcommand{\cCom}{\mathbin{\raisebox{0.5ex}{,}}}

\newcommand{\vphi}{{\text{\contour{black}{$\phi$}}}}
\newcommand{\vvphi}{{\text{\contour{black}{$\varphi$}}}}
\contourlength{0.022em}

\definecolor{tocColour}{rgb}{0.14902, 0.133333, 0.384314}

\begin{document}

\setcounter{page}0
\def\ppnumber{\vbox{\baselineskip14pt
}}

\def\ppdate{
} \date{\today}

\title{\LARGE \bf 
Collusion of Interactions and Disorder at the Superfluid-Insulator Transition:\\ A Dirty 2d Quantum Critical Point
}

\author[1,$\ddagger$]{Hart Goldman}
\author[2,3,$\ddagger$]{Alex Thomson}
\author[4]{Laimei Nie}
\author[5]{Zhen Bi}
\affil[1]{\it \small Department of Physics and Institute for Condensed Matter Theory,\protect\\  \it \small University of Illinois at Urbana-Champaign, Urbana, IL 61801, USA}
\affil[2]{\it \small Department of Physics and Institute for Quantum Information and Matter, \protect\\ \it \small California Institute of Technology, Pasadena, CA 91125, USA} 
\affil[3]{\it \small Walter Burke Institute for Theoretical Physics, \protect\\ \it \small  California Institute of Technology, Pasadena, California 91125, USA}
\affil[4]{\it \small Kadanoff Center for Theoretical Physics, University of Chicago,
Chicago, IL 60637, USA}
\affil[5]{\it \small Department of Physics, Massachusetts Institute of Technology, Cambridge, MA 02139, USA}
\date{}
\maketitle

\begin{abstract}
We study the stability of the Wilson-Fisher fixed point of the quantum $\mathrm{O}(2N)$ vector model to quenched disorder in the large-$N$ limit. While a random mass is strongly relevant at the Gaussian fixed point, its effect is screened by the strong interactions of the Wilson-Fisher fixed point. This enables a perturbative renormalization group study of the interplay of disorder and interactions about this fixed point. We show that, in contrast to the spiralling flows obtained in earlier double-$\epsilon$ expansions, 
the theory flows directly to a quantum critical point characterized by finite disorder and interactions. 
The critical exponents we obtain for this transition are in remarkable agreement with numerical studies of the superfluid-Mott glass transition. We additionally discuss the stability of this fixed point to scalar and vector potential disorder and use proposed boson-fermion dualities to make conjectures regarding the effects of weak disorder on dual Abelian Higgs and Chern-Simons-Dirac fermion theories when $N=1$.\vspace{0.15cm}\blfootnote{$^{\ddagger}$ These authors contributed equally to the development of this work.}   
\end{abstract}
\vspace{1cm}
\bigskip
\newpage

{
\hypersetup{linkcolor=tocColour}
\tableofcontents
}


\section{Introduction}
Many of the most challenging questions in condensed matter physics involve an interplay of quenched disorder and strong interactions in two spatial dimensions at zero temperature. 
A prominent example is the problem of understanding the nature of the field-tuned superconductor to insulator transition (SIT) in 
thin films. 
This transition not only appears to have the same critical exponents as the famously superuniversal quantum Hall plateau transitions \cite{Goldman1989,Jaeger1989,Paalanen1992,Steiner2005, Goldman2010,Breznay2017}, but also broadens into a finite metallic region in cleaner samples
\cite{Yazdani1995,Mason1999,Mason2001,2017arXiv170801908W,Kapitulnik2019}. 
Crucially, the universal data of this quantum phase transition has failed to appear in any theoretical construction involving disorder or interactions exclusively, indicating that both must play important roles.

\begin{figure}
\centering
\includegraphics[width=\textwidth]{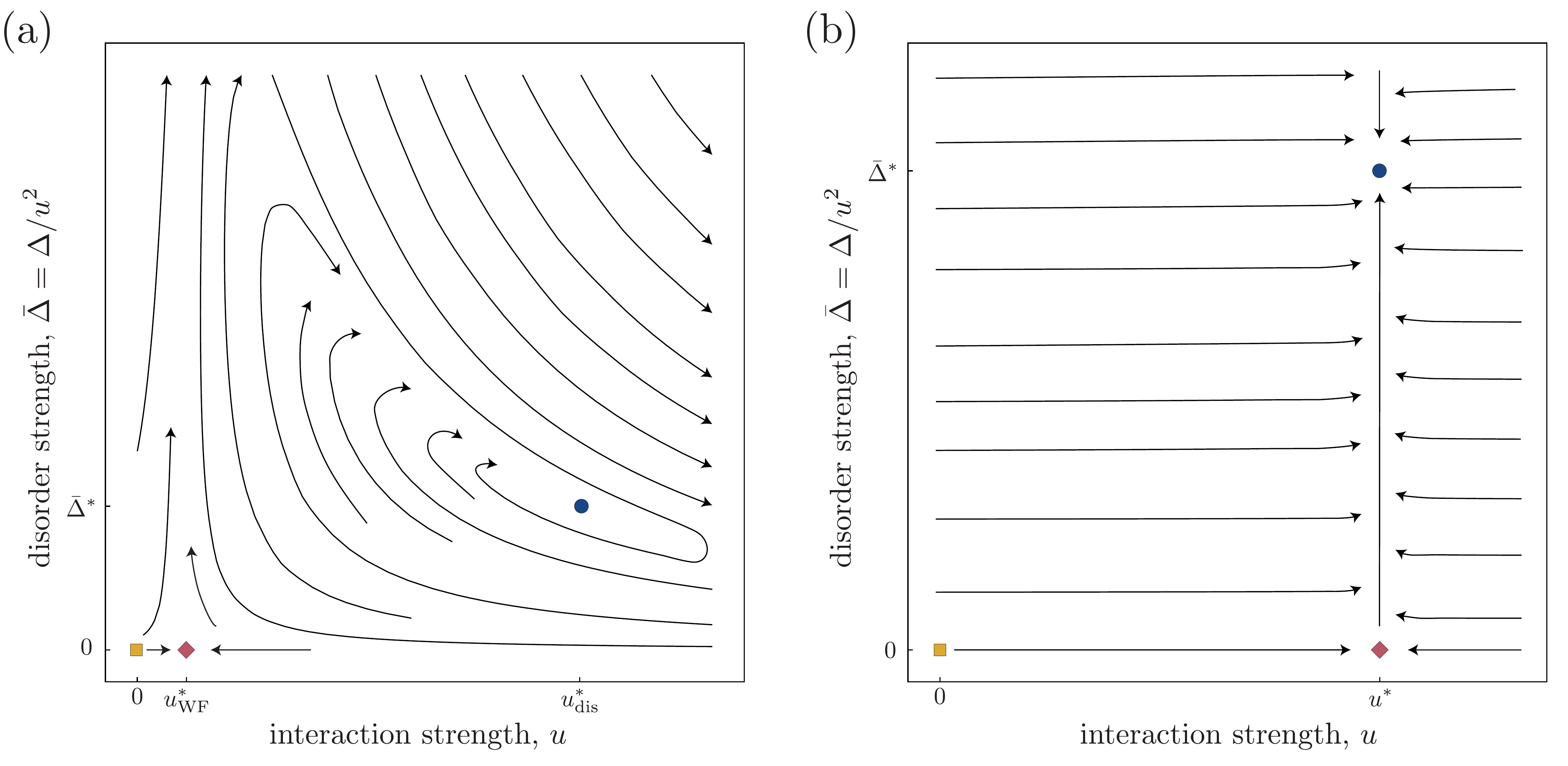}
\caption{
RG flow diagrams of the Gaussian fixed point (yellow square) as a function of the interaction coupling constant $u$ and running disorder strength $\bar{\Delta}=\Delta/u^2$.
The clean Wilson-Fisher fixed point is denoted by a red diamond, while the dirty fixed point is shown with a blue circle.
(a) Spiralling RG flows are obtained in the double-$\epsilon$ expansion for small numbers $N$ of bosons. (b) In the large-$N$ limit, we show that the Wilson-Fisher fixed point flows directly to a dirty, interacting quantum critical point. 
}
\label{fig:RGflows}
\end{figure}

In spite of decades of effort, few organizing principles have been developed for understanding quantum critical systems with interactions and disorder, and analytically tractable models have proven rare. This problem is particularly acute in bosonic systems undergoing SITs or superfluid-insulator transitions. While examples of quantum critical points and phases have been constructed in fermionic systems using perturbative and non-perturbative techniques \cite{Goswami2017,Thomson2017,Goldman2017,Maciejko2018}, few analogous examples exist for bosonic systems. At zero temperature, the only known examples of disordered-interacting fixed points of bosons in 2$d$ arise in the context of the superfluid-insulator transition of bosons with random mass disorder and $\phi^4$ interactions. These fixed points are obtained using a double-$\epsilon$ expansion about the free (Gaussian) fixed point in four spatial dimensions, perturbed with classical (finite temperature) disorder. 
This peculiar expansion, taken very far from the physical, quantum disordered situation of 
2+1 spacetime dimensions, was introduced by Dorogovtsev \cite{Dorogovtsev1980} and by Boyanovsky and Cardy \cite{Cardy1982}, who found a stable fixed point characterized by finite disorder and interactions (see also Ref. \cite{Lawrie1984}). 
However, the character of this fixed point is very strange and is not obviously of direct physical significance: the renormalization group (RG) flows in its vicinity are spirals. 
As Fig.~\ref{fig:RGflows}(a) demonstrates, it therefore takes a 
long time to approach this fixed point, and the critical regime may in fact be 
physically inaccessible. 
Fixed points with similar RG flows have been obtained in systems of bosons with $z=2$ \cite{Belitz1996} as well as in holographic constructions \cite{Hartnoll2014,Hartnoll2016}.

The view we take in this work is that the unusual character of the double-$\epsilon$ expansion fixed point 
may be understood as an artifact of perturbing the free, classical  
fixed point. 
Near such a fixed point, disorder can prematurely take control of the physics, obscuring the true fate of the strongly interacting, disordered theory. 
Indeed, the technical reason\footnote{See Refs. \citenum{Aharony2018} and~\citenum{Aharony2018a} for a more detailed discussion.} for the appearance of spiralling flows is that at the free, classical fixed point
, the $\phi^4$ operator and the operator associated with the quenched disorder (in the replica formalism) 
have the same scaling dimension.
As a result, these operators can immediately mix along the RG flow in such a way that 
their scaling dimensions enter the complex plane, leading to the spirals in Fig.~\ref{fig:RGflows}(a). 
In contrast, the appearance of complex scaling dimensions is not expected to occur near the Wilson-Fisher fixed point, where these operators do not have the same scaling dimensions. A hint that this is the case comes from studying the double-$\epsilon$ expansion RG equations in the limit of a large number $N$ of boson species. 
In this limit, the scaling dimensions of these operators near the Wilson-Fisher fixed point are far from degenerate, and there is a crossover into a regime in which complex scaling dimensions no longer occur. 

In this article, we demonstrate that the strongly coupled Wilson-Fisher fixed point gives way to a quantum critical point (QCP) distinguished by both finite disorder and interactions using a large-$N$ expansion. 
Instead of simultaneously perturbing the free, classical fixed point with both disorder and interactions, as in the double-$\epsilon$ expansion, 
we introduce weak disorder directly at the fully quantum, interacting Wilson-Fisher fixed point.  
While this fixed point saturates the Harris criterion in the $N\rightarrow\infty$ limit, we find that it is destabilized at $\mathcal{O}(1/N)$, 
resulting in flows of the type shown in Fig.~\ref{fig:RGflows}(b).
This fixed point is characterized by a correlation length exponent $\nu$ and a dynamical scaling exponent $z$ given at $\mathcal{O}(1/N)$ by
\be
\label{eq: intro exponents N}
\nu=1\,,\qquad z=1+\frac{16}{3\pi^2N}\cdot
\ee
Extrapolation to $N=1$ therefore yields a value $z\approx1.5$ for the $\mathrm{O}(2)$ model. The associated operator scaling dimensions are presented alongside the critical exponents of the clean fixed point in Table \ref{tab:ScalingDims}. 

The values these exponents take have several noteworthy implications.
The correlation length exponent $\nu$ at the disordered fixed point is the same as at the clean Wilson-Fisher fixed point in the large-$N$ limit.
This absence of $1/N$ corrections may be interpreted as a physical consequence of the balancing that occurs between disorder and interaction effects. 
On the other hand, the fact that $1<z<2$ signals that the fixed point is neither clean nor conventionally diffusive ($z=2$), a hallmark of dirty quantum criticality. A similar physical story occurs in the earlier studies of disorder in QED$_3$ \cite{Goswami2017,Thomson2017}.

{\renewcommand{\arraystretch}{2}
\begin{table}
\centering
\begin{tabular}{rclclcl}
&   &  \multicolumn{1}{c}{$z$} 
&   &  \multicolumn{1}{c}{$\big[\phi\big]$}  
&   &  \multicolumn{1}{c}{$\big[\phi^2\big]$} 
\\\hline\hline
disordered WF
&   &   $\displaystyle 1+{16\o3\pi^2N}$   
&   &   $\displaystyle {1\o2}+{2\o3\pi^2N}$    
&   &   $\displaystyle 2+{16\o3\pi^2N}$ 
\\
clean WF
&   &   $\displaystyle 1$
&   &   $\displaystyle {1\o2}+{2\o3\pi^2N}$
&   &   $\displaystyle 2-{16\o3\pi^2N}$
\end{tabular}
\caption{Scaling dimensions at the dirty, interacting QCP obtained in the large-$N$ expansion, compared with the results 
the clean Wilson-Fisher fixed point at large-$N$. Here $\phi$ denotes the boson field, and $\phi^2$ denotes the mass operator.
The correlation length exponent $\nu$ is obtained through $\nu^{-1}=2+z-\big[\phi^2\big]$.
}
\label{tab:ScalingDims}
\end{table}
}

The QCP we obtain   
may be relevant to superfluid-insulator transitions in $^4$He absorbed in porous Vycor \cite{Reppy1983,Reppy1984,Reppy1988}, Josephson junction arrays \cite{Mooij1992,Mooij1996}, doped quantum magnets \cite{Tanaka2002,Regnault2010,Yu2012}, and cold atomic systems \cite{White2009,Krinner2014,Modugno2014}. Superfluid-insulator transitions with similar exponents have also been observed numerically \cite{Prokofev2004,Priyadarshee2006,Pollet2009, Meier2012,Ng2015,Vojta2016}. 
Indeed, the values we obtain at $\mathcal{O}(1/N)$ for $\nu$, $z$, and the correlation function exponent $\eta\approx-0.47$ are strikingly close to those obtained in the most recent Monte Carlo study of the dirty $\mathrm{O}(2)$ model \cite{Vojta2016}. 
This achievement is all the more striking given that it comes from extrapolating $N$ to $1$. 
However, we note that in these numerical approaches the insulating phase is either a ``Mott glass,'' 
which is incompressible \cite{Giamarchi2001,Weichman2008}, or a ``Bose glass,''
which has finite compressibility. While it is generally believed that the superfluid state always gives way to a glassy insulator in 2$d$ \cite{Fisher1988,Fisher1989}, assessing whether this is the case in the theory examined here requires the inclusion of non-perturbative effects, which are beyond the scope of our discussion here. 

Similar large-$N$ approaches to the study of quenched disorder at the Wilson-Fisher fixed point have been applied in the past by Kim and Wen \cite{Kim1994} and by Hastings \cite{Hastings1999}. In the latter case, $1/N$ corrections were not considered, while in the former runaway flows were obtained. We believe that these runaway flows are the result of a redundant summation of diagrams. 


We proceed as follows. In Sec.~\ref{sec:StabilityCriterion}, we present a stability criterion for theories of interacting bosons to quenched disorder. 
We next perform the large-$N$ analysis and describe the nature of the QCP we obtain in Sec.~\ref{sec:RG}.
We follow in Sec.~\ref{sec:CurrentDisorder} with a discussion of the effects of scale and vector potential disorder.
In Sec.~\ref{sec:Duality}, the implications our result has for two dual descriptions of the single species ($N=1$) theory, following earlier work coauthored by one of us \cite{Goldman2017}.
We conclude in Sec.~\ref{sec: discussion}.

\section{Stability Criterion for Free and Interacting Bosons}
\label{sec:StabilityCriterion}

We begin this section by describing the criteria for the stability of theories of relativistic scalar bosons to quenched disorder at zero temperature, often referred to as quantum disorder.
After presenting our conventions and the global symmetries, we derive a criterion for the free, Gaussian fixed point.
We then generalize this criterion to the strongly interacting, Wilson-Fisher fixed point, where anomalous scaling dimensions appear. These stability criteria are quantum bosonic versions of the celebrated Harris criterion \cite{Harris1974}.    

\subsection{Degrees of Freedom and Global Symmetries}

We consider one of the simplest families of quantum field theories: those describing massless, complex scalar fields transforming in the fundamental representation of $\mathrm{U}(N)$. 
Writing the bosonic degrees of freedom as $N$-component complex vectors $\vphi=(\phi_1,\dots,\phi_N)$, this global symmetry acts as $\vphi\to U\vphi$, $U\in\mathrm{U}(N)$. 
Throughout this paper, we restrict our attention to disorder and interactions that respect the full $\mathrm{U}(N)$ symmetry. 

For the majority of this work,
we also impose two additional discrete, anti-unitary symmetries:
time reversal, $\T$, and particle-hole symmetry, $\PH$. 
They act on the fields as
\begin{align}
\T&:\,\vphi\mapsto\vphi\,,\\
\PH&:\,\vphi\mapsto\vphi^\dagger\,,
\end{align}
and both map $i\mapsto-i$.
We eventually consider types of disorder that break these within each realization while preserving them on average in Sec.~\ref{sec:CurrentDisorder}.

When the above global symmetries are imposed, the theory of $\vphi$ fields is also invariant under the larger symmetry group, $\mathrm{O}(2N)$. 
Its action is obtained by defining $2N$ real fields, $\varphi_I$, from the complex fields: $\phi_I=\varphi_{2I-1}+i\varphi_{2I}$.
The theory we discuss below is found to be invariant under the action of $\vvphi\to O\vvphi$ where $O\in \mathrm{O}(2N)$ and $\vvphi=(\varphi_1,\dots,\varphi_{2N})$.
Actually, the orthogonal global symmetry need not only arise as an enhanced symmetry, but can exist as a true global symmetry even away from the critical point. 
For such cases, there is no reason to the restrict the number of flavors to be even.
Hence, while we primarily discuss the complex fields $\vphi$,  we allow $N$ to take half-integer values.




\subsection{Free Bosons with Disorder}
We begin with a free, or Gaussian, theory of $N$ complex bosons,
\eq{
\mathcal{L}_0[\vphi]=\abs{\pd\vphi}^2,
}
in $d+1$ spacetime dimensions. 
Throughout this paper, `$d$' exclusively denotes the spatial dimension.
Dimensional analysis sets the scaling dimension of $\vphi$ to $[\vphi]=(d-1)/2$, and the scaling of all operators in the free theory follows directly from this relation. 
The stability of the Gaussian theory is determined 
by assessing the relevance of all 
operators respecting the global symmetries
described above.
The most relevant such perturbation is the mass term, 
 $r\abs{\vphi}^2$, 
since $[r]=2$ for all dimensions, and
the requirement that the theory be massless is therefore predicated on the fine-tuning of $r$ to zero.
The next-most relevant, symmetry-preserving operator is the interaction term $u\abs{\vphi}^4=u\left(\abs{\vphi}^2\right)^2$. 
Because $[\abs{\vphi}^4]=2(d-1)$, we have $[u]=3-d$, implying that the Gaussian theory becomes unstable to this interaction when $d<3$.
In the next section, we discuss the effect of adding this term.

Disorder is introduced by perturbing $\mathcal{L}_0$ with an operator whose coefficient is a spatially varying, static field with values drawn from a probability distribution. 
Similar to the clean case, the most relevant, symmetry-preserving perturbation couples to the mass operator $\abs{\vphi}^2$:
\be
\mathcal{L}_0[\vphi, R]=|\pd\vphi|^2+R(\mathbf{x})|\vphi|^2,
\ee
where bold face denotes purely spatial coordinates.
We define $R(\mathbf{x})$ to have moments,
\be
\label{eq: mass disorder moments}
\overline{R(\mathbf{x})R(\mathbf{0})}\sim\frac{\Delta}{|\mathbf{x}|^\chi}\,,\qquad
\overline{R(\mathbf{x})}=0\,.
\ee
where $\chi\rightarrow d$ corresponds to Gaussian white noise\footnote{More precisely, one writes the disorder correlations as a Riesz potential, \begin{equation*}
\overline{R(\mathbf{x})R(\mathbf{0})}=\frac{\Gamma\left(\chi\over2\right)}{2^{d-\chi}\pi^{d/2}\Gamma\left({d-\chi}\over2\right)}\frac{\Delta}{|\mathbf{x}|^\chi}\,.
\end{equation*} 
It is this function that reproduces Gaussian white noise (delta function) correlations in the limit $\chi\rightarrow d$. 
In this paper, we will generally suppress the additional gamma functions, as these do not impact scaling.}. 
As it couples to $|\vphi|^2$, the dimension of $R(\vb{x})$ is $2$, just like the constant mass coefficient, $r$. 
From Eq.~\eqref{eq: mass disorder moments}, it follows that the engineering dimension of the disorder strength $\Delta$ at the Gaussian fixed point is
\be
[\Delta]=4-\chi\,.
\ee
For Gaussian white noise disorder, $\chi\rightarrow d$, implying that the theory is stable to random mass disorder provided that
\be
d>4\,,
\ee
which is the Harris criterion for free (relativistic) scalar fields. 

Comparing against our brief analysis of the clean theory, we observe that mass disorder is marginal when $d=4$,  whereas the $\abs{\vphi}^4$ interaction term is marginal when $d=3$.
This mismatch between the marginal dimensions associated with the disorder and interactions has been one of the major sources of difficulty in studying the dirty boson problem in two dimensions. 

We note that while the disorder perturbation $R(\vb{x})\abs{\vphi}^2$ and interaction term $\abs{\vphi}^4$ were chosen as the most relevant operators preserving the $\mathrm{U}(N)$, $\T$, and $\PH$ symmetries, they are also invariant under the $\mathrm{O}(2N)$ symmetry discussed in the previous section. 
When the discrete symmetries, $\T$ and $\PH$, are no longer imposed, additional $\mathrm{O}(2N)$-breaking perturbations are allowed.
We leave this discussion to Sec.~\ref{sec:CurrentDisorder}.


\subsection{Wilson-Fisher Bosons with Disorder}

When $d<3$, the Gaussian fixed point is unstable to \emph{both} disorder and $|\vphi|^4$ interactions. In the clean limit, this leads to the famous Wilson-Fisher 
fixed point,
\be
\mathcal{L}[\vphi]=|\pd\vphi|^2+r_c\,|\vphi|^2+\frac{u}{2N}\,|\vphi|^4\,.
\ee
Here $u=\Lambda^{3-d}\,\bar{u}$, $\bar{u}\sim\mathcal{O}(1)$, where $\Lambda$ is a UV cutoff scale. 
The mass $r_c$ tunes the theory to criticality. 
Its exact value is not physically meaningful, and we set it to zero throughout this work. 
At the Wilson-Fisher fixed point, the dimension of $|\vphi|^2$ differs from its engineering dimension (\emph{i.e.} scaling dimension in the free theory) by an anomalous dimension  $\eta_{|\phi|^2}$,
\be
\left\langle|\vphi(x)|^2|\vphi(0)|^2\right\rangle\sim\frac{1}{|x|^{2(d-1+\eta_{|\phi|^2})}}.
\ee
That is, the scaling dimension of $\abs{\vphi}^2$ is $[|\vphi|^2]=d-1+\eta_{|\phi|^2}$. 
Importantly, the anomalous dimension $\eta_{\abs{\phi}^2}$ is  afunction of the number of fields (and hence the symmetry of the theory).

We now perturb this fixed point with disorder,
\be
\mathcal{L}[\vphi, R]=|\pd\vphi|^2+R(\mathbf{x})|\vphi|^2+\frac{u}{2N}|\vphi|^4,
\label{eq:WF+RandomMass}
\ee
where $R(\mathbf{x})$ continues to be defined as in Eq.~\eqref{eq: mass disorder moments}. 
The dimension of $R$ is related to the scaling dimension of $|\vphi|^2$ as follows, 
\be
 \left[|\vphi|^2\right]=d-1+\eta_{|\phi|^2}=d+1-[R]\,.
\ee
With Eq.~\eqref{eq: mass disorder moments}, we can now read off the scaling dimension of the disorder strength:
\be
[\Delta]=2[R]-\chi=4-2\eta_{|\phi|^2}-\chi\,.
\ee
We conclude that the Wilson-Fisher fixed point is stable to Gaussian white noise disorder ($\chi\rightarrow d$) if
\be
d>4-2\eta_{|\phi|^2}\,.
\ee

\subsection{\texorpdfstring{Large-$N$ Wilson-Fisher in $(2+1)d$}{Large-N Wilson-Fisher in 2d}}
We now adapt this discussion to the particular case of a theory of $N\rightarrow\infty$ species of complex bosons in $d=2$ spatial dimensions.
In this limit, the stability criterion derived above becomes,
\eq{
\eta_{\abs{\phi}^2}-1>0.
}
For a single species of complex boson, it is known from the conformal bootstrap that ${\eta_{|\phi|^2}\sim0.5}$ in 2$d$ \cite{Kos2014}, implying that disorder is a relevant perturbation when $N=1$.
Conversely, in the limit $N\rightarrow\infty$, with $u$ held fixed, it turns out that $\eta_{|\phi|^2}\rightarrow 1$, as we will review in the next section. 
As a result, Gaussian white noise disorder ($\chi=2$) is \emph{marginal} at the Wilson-Fisher fixed point in the large-$N$ limit! The interacting dirty boson problem can therefore be studied by first flowing to the $N\rightarrow\infty$ Wilson-Fisher fixed point and subsequently performing a perturbative RG calculation, with $1/N$ corrections entering as marginal perturbations of the $N\rightarrow\infty$ fixed point. This will be the goal of the next section. 

Interpolating between the $N=1$ limit, where $\eta_{\abs{\phi}^2}\sim0.5$, and the $N\to\infty$ limit, where $\eta_{\abs{\phi}^2}\to1$, we expect $1/N$ corrections to $[|\vphi|^2]$ to be \emph{negative}, indicating that the Wilson-Fisher fixed point is ultimately unstable to disorder for finite $N$. 
Nevertheless, disorder generates additional corrections to scaling dimensions as well.
Provided these quantum corrections to $[\abs{\vphi}^2]$ are \emph{positive}, they may be able to balance the corrections from interactions, thus resulting in a perturbatively accessible, disordered quantum critical point.
In contrast, if the quantum corrections due to disorder are also negative, no such fixed point can exist, and all perturbations result in a flow to strong disorder. 
Serendipitously, we find that it is the former scenario which is played out.


\section{\texorpdfstring{The $\mathrm{O}(2N)$ Model with a Random Mass}{The O(2N) Model with a Random Mass}}\label{sec:RG}

This section presents the primary technical content of the paper. 
We begin by describing the disorder-averaged theory and its replicated analogue.
Next, the number of bosons $N$ is taken to infinity, leaving us with a theory in which disorder is exactly marginal.
We subsequently derive the $\b$ function for the running disorder strength at $\mathcal{O}(1/N)$ and demonstrate the existence of the fixed point and RG flow shown in Fig.~\ref{fig:RGflows}(b).
The section concludes with a comparison of the fixed point obtained here with the results from the double-$\epsilon$ expansion.

\subsection{Disorder Averaging and the Replica Trick}
We now describe how to systematically study the dirty Lagrangian in Eq.~\eqref{eq:WF+RandomMass} in the ${N\to\infty}$ limit.
While the addition of the quenched degree of freedom $R(\mathbf{x})$ strongly breaks translation invariance, seemingly rendering the theory intractable, we are interested in the disorder-averaged correlation functions, for which translation symmetry remains. 
In direct correspondence with the clean case, all quantities of interest in the disordered theory may be calculated from the disorder-averaged free energy:
\begin{align}
\bar{F}
&=
-\overline{\log Z[R]}
=
-\int\mathcal{D}R\,\mathcal{P}[R]\,\log Z[R]\,,
\end{align}
where $\mathcal{P}[R]$ is the probability distribution that gives rise to the moments in Eq.~\eqref{eq: mass disorder moments}. 
Specifying to Gaussian white noise disorder, the appropriate probability functional is
\be
\mathcal{P}[R]=\frac{1}{\mathcal{N}}\,\exp\left(-\int d^2\mathbf{x}\,{1\over{2\Delta}}R^2(\mathbf{x})\right),
\ee 
where $\mathcal{N}$ is a normalization constant.

While directly disorder averaging the logarithm is essentially futile, we appeal to the so-called replica trick, a method founded on the application of the identity
\be
\log Z=\lim_{n_r\rightarrow0}\frac{Z^{n_r}-1}{n_r}\cdot
\ee
Upon inserting this expression into the definition of $\bar{F}$, we obtain
\begin{align}
\bar{F}&=-\lim_{n_r\rightarrow0}\frac{1}{n_r}\int\mathcal{D}R\,\mathcal{P}[R]\prod_{n=1}^{n_r}\int\,\mathcal{D}\vphi_{n}\,e^{-S[\vphi_{n},R]}
\notag\\
&
=-\lim_{n_r\rightarrow0}\int\mathcal{D}R\,\mathcal{D}\vphi_{n}\,e^{-S_{r}[\vphi_{n},R]}\,,
\end{align}
where $S=\int d^2\mathbf{x}\,d\tau\,\mathcal{L}[\vphi_{n},R]$ and $n_r$ ``replicas," $\vphi_{n}$, $n=1,\dots,n_r$, have been introduced. 
We remind the reader that each replica is associated with $N$ physical species of bosons. 
 The full replicated action for Gaussian white noise disorder is 
\begin{align}
S_{r} &=  \int d^2\mathbf{x}\,d\tau\sum_{n=1}^{n_r}\left[|\partial\vphi_{n}(\mathbf{x},\tau)|^2+ \frac{R(\mathbf{x})}{\sqrt{N}} |\vphi_{n}(\mathbf{x},\tau)|^2 + \frac{u}{2N} \,|\vphi_{n}(\mathbf{x},\tau)|^4 \right] \nonumber\\
&\quad +\int d^2\mathbf{x}\,\frac{1}{2\Delta}R^2(\mathbf{x}).
\end{align}
Here, $R$ has been rescaled by $\sqrt{N}$, equivalent to rescaling $\Delta$ by $1/N$.
In summary, the replica trick has produced an action amenable to the standard tools of perturbative field theory through the addition of $n_r$ replica fields, with the caveat that we must eventually take the unsavory limit $n_r\rightarrow0$. 


\subsection{\texorpdfstring{The Large-$N$ Limit}{The Large-N Limit}}

Fixing the value of $\Delta$ and $u$, we are now able to take the large-$N$ limit.
It is convenient to introduce a Hubbard-Stratonovich field $i\tilde \sigma$ (the reason for the tilde will become apparent shortly) to mediate the scalar self-interaction:
\begin{align}
S_{r}=\int d^2\mathbf{x}\,d\tau \sum_n\left[ 
|\pd\vphi_{n}|^2
+ 
\frac{1}{\sqrt{N}}\Big( i\tilde\sigma_n+  {R(\mathbf{x})} \Big) |\vphi_{n}|^2
+  
\frac{1}{2u} \tilde\sigma_n^2 
\right]
+\int d^2\mathbf{x}\,\frac{1}{2\Delta}R^2(\mathbf{x})\,.
\end{align}
The equations of motion for $i\tilde{\sigma}$ directly relate it to the mass operator
\be\label{eq:OperatorIdentity}
i\tilde\sigma_n=\frac{u}{\sqrt{N}}\,|\vphi_{n}|^2\,,  
\ee
and it follows that correlation functions containing $i\tilde\sigma$ will reproduce correlation functions containing $|\vphi|^2$ up to an contact term. 
Next, we shift $i\tilde{\sigma}_n\to i\sigma_n= i\tilde{\sigma}_n+R$ so that the coupling between $R$ and the $\vphi$ fields is replaced with a coupling between $R$ and $\sigma$, 
\be\label{eqn:RepActionR}
S_{r}=\int d^2\mathbf{x}\,d\tau \sum_n \left[|\pd\vphi_{n}|^2+\frac{i}{\sqrt{N}} \,\sigma_n|\vphi_{n}|^2 + \frac{i}{u}\, R(\mathbf{x})\,\sigma_n +\frac{1}{2u}\sigma_n^2 \right]
+\int d^2\mathbf{x}\frac{1}{2\Delta}R^2(\mathbf{x}).
\ee
Here, an extra term quadratic in $R$ is not included because it is proportional to the number of replicas and therefore vanishes in the replica limit. 
Finally, integrating out the quenched degree of freedom $R(\mathbf{x})$  yields
\begin{align}
S_{r}&=\int d^2\mathbf{x}\,d\tau \sum_n\left[|\pd\vphi_{n}|^2+\frac{i}{\sqrt{N}}\,\sigma_n|\vphi_{n}|^2 + \frac{1}{2u}\sigma_n^2\right]\nonumber\\
&\quad+\int d^2\mathbf{x}\,d\tau d\tau' \sum_{n,m}\frac{\Delta}{2u^2}\,\sigma_n(\mathbf{x},\tau)\sigma_m(\mathbf{x},\tau')\,.
\label{eq: replicated bare action}
\end{align}
Equipped with this Lagrangian, we are now prepared to take the large-$N$ limit following the standard procedure. 
For a more detailed review, see Refs.~\citenum{Polyakov1987} and~\citenum{Moshe2003}.

\begin{figure}
\includegraphics[width=\textwidth,page=1]{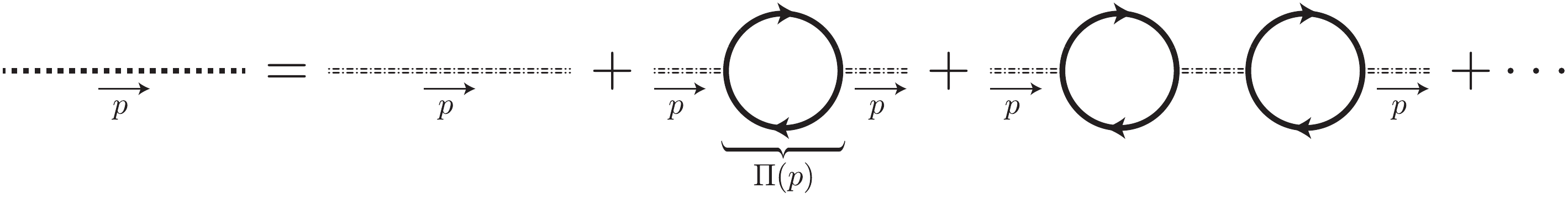}
\caption{In the $N\rightarrow\infty$ limit, the propagator of $\sigma$ may be represented as a geometric series of polarization bubbles $\Pi(p)$.
The dash-dotted lines on the right-hand side represent the `bare' $\s$ propagator $\sim1/u$, whereas the solid lines represent the $\phi$ propagator. 
The dotted line corresponds to the large-$N$ $\s$ Green's function.
}
\label{fig: large-N}
\end{figure}
We begin by noting that the action $S_r$ is quadratic with the exception of the $\sigma|\vphi|^2$ interaction.
While $\sigma$ couples more and more weakly to the $\vphi$'s as $N$ approaches infinity, it also couples to increasingly many such fields.
The result of these opposing effects can be understood in the language of Feynman diagrams. 
In particular, the one-loop contribution to the  $\sigma$ propagator is the polarization bubble shown in Fig.~\ref{fig: large-N}.
Because the internal boson lines must be summed over all $N$ fields while each vertex contributes a factor of $1/\sqrt{N}$, this diagram is $\mathcal{O}(1)$.
It evaluates to
\be
\Pi(p)=\int\frac{d^3k}{(2\pi)^3}\frac{1}{k^2(p-k)^2}=\frac{1}{8|p|}\cdot
\ee
Of course, if a diagram containing a single bubble is $\mathcal{O}(1)$, a diagram containing an arbitrary number of bubbles is also $\mathcal{O}(1)$, and so it should be include as well. 
The sum over bubble diagrams forms the geometric series shown in Fig.~\ref{fig: large-N}, which may be familiar to readers trained in the random phase approximation. 
The large-$N$ $\sigma$ propagator is therefore 
\be
G^\sigma(p)=\frac{u}{1+u\,\Pi(p)}\rightarrow 8|p|\qquad \text{for }p \ll u\,, 
\label{eq: sigma prop 1}
\ee
where we have taken $u\sim\Lambda$ as our UV cutoff. 

The physical meaning of these bubble diagrams can be understood by considering the real space representation of $G^\sigma$, which has been ``screened'' to be short-ranged,
\be
G^\sigma(x)=\langle\sigma(x)\sigma(0)\rangle\sim\frac{1}{|x|^4}\cdot
\ee
The large-$N$ $\sigma$ propagator makes 
it clear that $[\sigma]=2$ when $N\to\infty$, implying that $\sigma$ has acquired an anomalous dimension $\eta_\sigma=\eta_{|\phi|^2}=1$, as claimed in the previous section.


Having accounted for the effect of bubble diagrams, the interaction between $\vphi$ and $\sigma$ may be safely discarded in the limit $N\rightarrow\infty$.
It is possible to access $1/N$ corrections by reintroducing the coupling between $\vphi$ and $\sigma$ and using the screened  $\sigma$ propagator in Eq.~\eqref{eq: sigma prop 1} on the condition that bubble diagrams are not redundantly included in any subsequent calculation. 
Keeping this is mind, we obtain the effective action
\bea
\label{eq: large N effective action}
S_{\textrm{eff}}&=&S_{\phi}+S_{\sigma\phi}+S_{\mathrm{dis}}\\
\label{eq: phi kinetic energy}
S_{\phi}&=&\sum_{n}\int d^2\mathbf{x}\,d\tau\,|\pd\vphi_{n}|^2\\
S_{\sigma\phi}&=&\sum_n\int d^2\mathbf{x}\,d\tau\left[\frac{i}{\sqrt{N}}\,\sigma_n|\vphi_{n}|^2+\frac{1}{16}\,\sigma_n(-\pd^2)^{-1/2}\sigma_n\right] \\
\label{eq: disorder action}
S_{\mathrm{dis}}&=&\sum_{n,m}\int d^2\mathbf{x}\,d\tau d\tau'\,\frac{\bar\Delta}{2}\,\sigma_n(\mathbf{x},\tau)\,\sigma_m(\mathbf{x},\tau')\,,
\eea
where we have defined the dimensionless disorder strength $\bar\Delta\equiv\Delta/u^2$. 

We are interested in the effect nonzero $\bar{\Delta}$ has on 
this theory, which we emphasize is now a \emph{marginal} perturbation at tree level. 
Indeed, the disorder-mediated potential between two $\vphi$ fields has been screened to be
\be
V(\mathbf{x}-\mathbf{y})\sim\frac{\bar\Delta}{|\mathbf{x}-\mathbf{y}|^4}\cdot
\ee



\subsection{\texorpdfstring{$1/N$ Corrections: Introducing Disorder at the Interacting Fixed Point}{1/N Corrections}}

\subsubsection{Philosophy and Scaling Conventions}\label{sec:ThePlan}

We include the effects of disorder and interactions at $\mathcal{O}(1/N)$ via a Wilsonian momentum shell RG procedure.
To begin, we present our tree-level scaling conventions.
The action in Eq.~\eqref{eq: large N effective action}, including the disorder, is scale invariant under 
\be
\mathbf{x} \mapsto e^{\delta\ell} \mathbf{x}, 
\qquad
\tau \mapsto e^{z\delta\ell} \tau,
\qquad
\vphi \mapsto e^{-\delta\ell/2} \vphi,
\qquad
\sigma\mapsto e^{-2\delta\ell}\sigma\,.
\label{eq: rescaling}
\ee
Lorentz invariance dictates that space and time scale in the same way at the clean Wilson-Fisher fixed point; hence, $z=1$.
The scaling 
prescriptions
for $\vphi$ and $\sigma$ are in agreement with our earlier statement that $[\vphi]=1/2$ and $[\sigma]=2$ in the $N\to\infty$ limit of the Wilson-Fisher fixed point.
At $\mathcal{O}(1/N)$, these relations must be updated to account for anomalous dimensions generated by disorder and interactions, which we denote $\eta_\phi$ and $\eta_\sigma$ for the $\vphi$ and $\sigma$ fields, respectively.
Similarly, because disorder breaks Lorentz invariance, the dynamical exponent is corrected to a value $z>1$. 
We systematically compute these corrections to scaling by integrating out modes in a frequency shell $(1-\delta\ell)\Lambda < |\mathbf{p}| < \Lambda$, where $\Lambda \sim u$ is a hard cutoff. 
Note that because of the large-$N$ limit, we may take $\bar\Delta\sim\mathcal{O}(1)$, as our perturbation theory continues to be controlled in powers of $1/N$. 

Before presenting the details of our calculation, we remark on some idiosyncrasies of the theory \eqref{eq: large N effective action} that ultimately serve to simplify our analysis. 
We first comment on the clean limit, $\bar\Delta=0$. 
Quantum corrections are typically organized into self energy corrections and vertex corrections, which modify the scaling of the fields and affect the running of the interactions. 
In the theory \eqref{eq: large N effective action}, we would therefore expect $\sigma|\vphi|^2$ to enter in the Lagrangian alongside a running coupling constant. 
However, because $\sigma$ was defined through a Hubbard-Stratonovich transformation, it is not independent from $|\vphi|^2$, as indicated by the operator identity of Eq.~\eqref{eq:OperatorIdentity}. 
It follows that the $\sigma|\vphi|^2$ vertex remains exactly marginal under the RG, making the renormalization of this vertex sufficient to determine $\eta_\sigma$, the anomalous dimension of $\sigma$. 
This observation is advantageous because the corrections to $\sigma|\vphi|^2$  all occur at one loop, whereas a direct calculation of the $\sigma$ self energy involves the computation of two loop diagrams.

The introduction of disorder results in both a running disorder strength $\bar\Delta$ and the aforementioned dynamical scaling exponent $z$. 
It turns out that these are the only additional objects to be renormalized in our problem at $\mathcal{O}(1/N)$.
Further, we find that the running of both may be obtained solely through the $\vphi$ self energy and the $\sigma|\vphi|^2$ vertex correction, similar to the clean case discussed above.
The key consequence of this assertion is
that under the modified scaling relations $\tau\mapsto e^{z\delta\ell} \tau ,\mathbf{x}\mapsto e^{\delta\ell}\mathbf{x}$, and $\sigma\mapsto e^{-(2+\eta_\sigma)\delta\ell} \sigma$, 
\be
\beta_{\bar \Delta} = -\frac{\delta \bar \Delta}{\delta\ell} = 2\big( 1-z + \eta_\sigma   \big) \bar \Delta\,. 
\label{eq: beta function of disorder}
\ee
The remainder of the section is dedicated to the calculation of $z$ and $\eta_\sigma$.

We emphasize that this simplification is not a generic feature of the problem.
It is possible for 
logarithmically divergent diagrams to generate operators containing\footnote{For a more general discussion of this point, see Refs.~\citenum{Aharony2018} and~\citenum{Aharony2018a}.} $\sum_n\int d\tau\,\sigma_n(\mathbf{x},\tau)$ independently from $\sigma_n(\mathbf{x},\tau)$. 
Such mixing would invalidate Eq.~\eqref{eq: beta function of disorder}, as well as contribute to a running velocity for $\sigma$.
For this reason, the $\sigma$ self energy must also be computed. 
These considerations are reflected by the modification of Eq.~\eqref{eq:OperatorIdentity} in the presence of disorder, which now involves this new, linearly independent operator,
\be
i\sigma_n=\frac{u}{\sqrt{N}}|\vphi_n|^2-iu\,\bar\Delta\sum_m\int d\tau\,\sigma_m(\mathbf{x},\tau)\,.
\ee
We evaluate the $\sigma$ self energy in Appendix~\ref{app:DimReg} using a dimensional regularization scheme, a more natural method for higher loop calculations.
This calculation confirms that no such diagrams occur at $\mathcal{O}(1/N)$, although they may appear at higher orders.

\subsubsection{Feynman Rules}
\begin{figure}[t]
\includegraphics[width=0.9\textwidth,page=2]{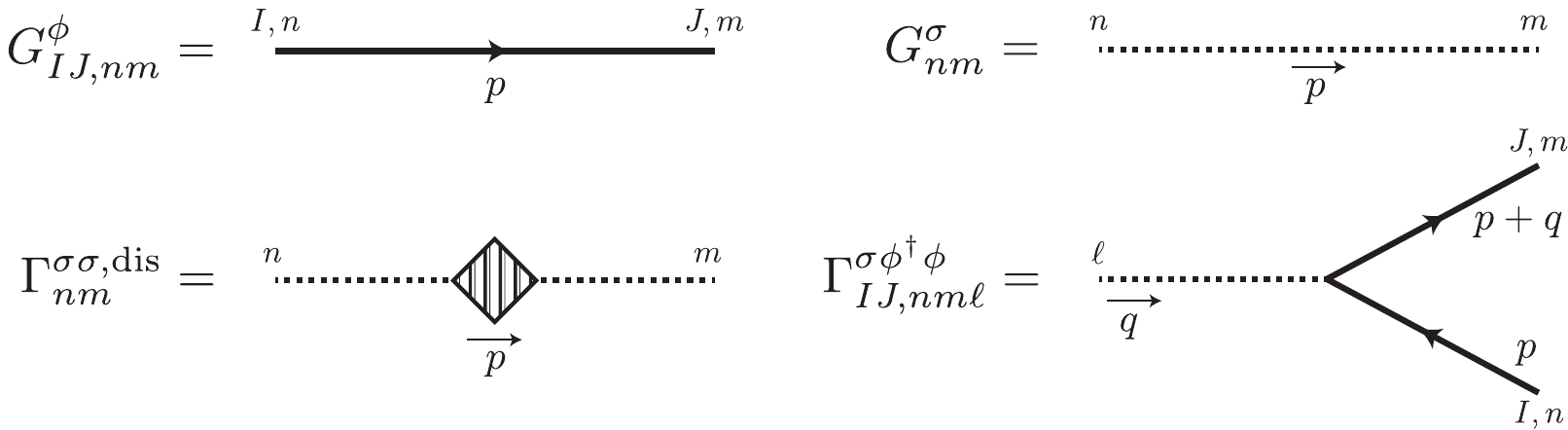}
\caption{Feynman rules for the theory \eqref{eq: large N effective action}.
Here, $p=(\vp,\omega)$, $q=(\vq,\nu)$, where $\vp,\vq$ are spatial momenta and $\omega,\nu$ are frequencies.
}
\label{fig: Feynman Rules}
\end{figure}

The Feynman rules for the theory in Eq.~\eqref{eq: large N effective action} are shown in Fig.~\ref{fig: Feynman Rules}, where
\begin{align}\label{eqn:FeynRules}
G_{IJ,nm}^\phi(p)&=\frac{1}{p^2}\,\delta_{IJ}\,\delta_{mn}\,,\\
G_{nm}^\sigma(p)&=8|p|\,\delta_{mn}\,,
\\
\Gamma^{\s\phi^\dag\phi}_{IJ,nm\ell}
&=-\frac{i}{\sqrt{N}}\,\delta_{IJ}\,\delta_{mn}\delta_{n\ell}\,,
\\
\Gamma^{\s\s,\mathrm{dis}}_{nm}
&=
-2\pi\bar\Delta\,\delta(\omega)\,.
\end{align}
Here, we have suppressed the momenta-conserving delta functions and use $I,J=1,\dots,N$ to denote flavor indices.  
Below, we suppress the $\mathrm{U}(N)$ and replica indices in the three-point vertex functions: $\Gamma^{\sigma\phi^\dag\phi}_{IJ,nm\ell}=\Gamma^{\s\phi^\dag\phi}$.
We also  emphasize that the quenched disorder is capable of transferring momentum, but not frequency, as indicated with the frequency of $\delta$-function.

We remark that disorder is being treated as a two-point vertex even though it appears as a quadratic field term in the action. 
While such terms are typically incorporated directly into the propagator, 
in our problem
$\sigma$ lines with multiple disorder insertions necessarily vanish in the replica limit, leaving only the contribution from the two-point vertex. 
We underscore
that this is a non-perturbative statement, as $\bar{\Delta}\sim\mathcal{O}(1)$.


\subsubsection{Momentum Shell RG} 

\begin{figure}
\begin{center}
\includegraphics[width=0.9\textwidth,page=3]{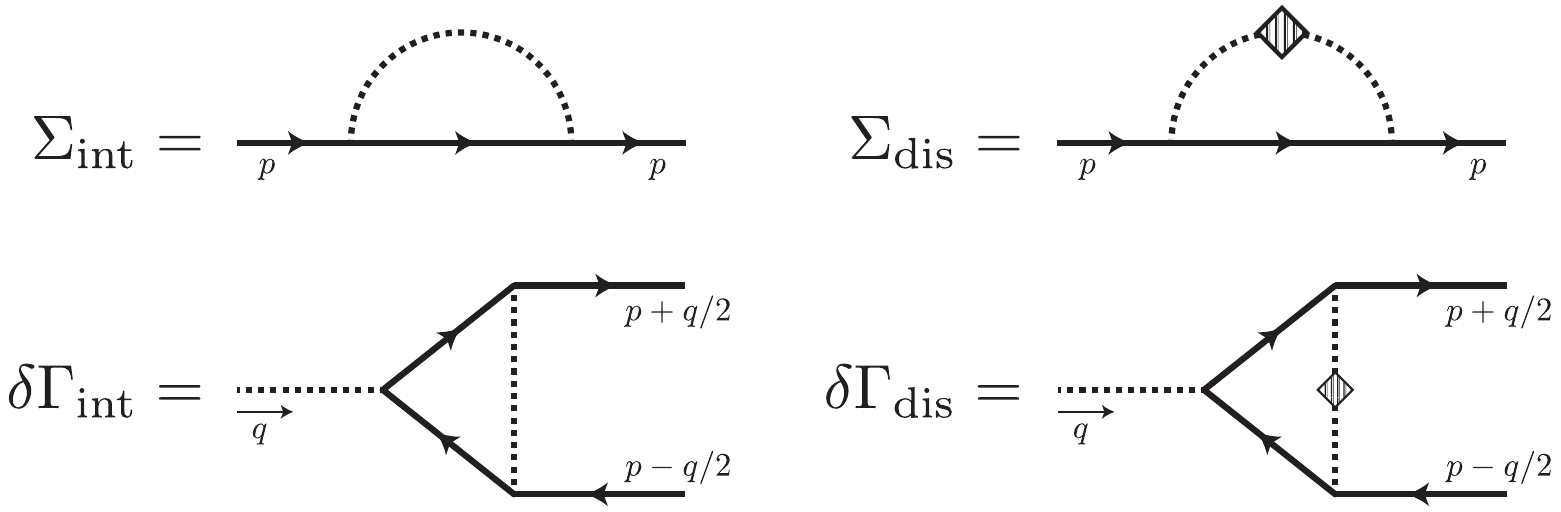}
\end{center}
\caption{Quantum corrections at $\mathcal{O}(1/N)$. (Top) $\phi$ self-energy corrections, $\Sigma_{\mathrm{int}}$ (left) and $\Sigma_{\mathrm{dis}}$ (right). (Bottom) Logarithmically divergent vertex corrections, $\delta\Gamma_{\mathrm{int}}$ (left) and $\delta\Gamma_{\mathrm{dis}}$ (right).
The full set of $\mathcal{O}(1/N)$ diagrams are shown in Fig.~\ref{fig:3ptVertex} in Appendix~\ref{app:DimReg}.
}
\label{fig: 1/N diagrams}
\end{figure}

We first focus on the $\phi$ self energy, as shown in Fig.~\ref{fig: 1/N diagrams}.  
After the momentum shell integration, we obtain 
\begin{align}
\Sigma(\mathbf{p},\omega) &= \Sigma_{\mathrm{int}} (\mathbf{p},\omega) + \Sigma_{\mathrm{dis}} (\mathbf{p},\omega),\\
\label{eq: SE int}
\Sigma_{\mathrm{int}} (\mathbf{p},\omega) &= -\frac{8}{N} \int_{(1-\delta\ell )\Lambda}^\Lambda \frac{d^2 \mathbf{k}}{(2\pi)^2}\int_{-\infty}^{\infty}\frac{d k_0}{2\pi} \frac{|k-p|}{k^2} = -\frac{4}{3 \pi^2 N} p^2 \delta\ell ,
\\
\label{eq: SE dis}
\Sigma_{\mathrm{dis}} ( \mathbf{p},\omega)
&=
\frac{64 \bar \Delta}{N} \int_{(1-\delta\ell )\Lambda}^\Lambda \frac{d^2 \mathbf{k}}{(2\pi)^2} \frac{(\mathbf{k} - \mathbf{p})^2}{\omega^2 + |\mathbf{k}|^2} = \frac{32 \bar \Delta}{ \pi N} (-\omega^2 + |\mathbf{p}|^2)  \delta\ell.
\end{align}
These correct the kinetic term of $S_\phi$, Eq.~\eqref{eq: phi kinetic energy}; the mass renormalization has been suppressed. 
To maintain the scale invariance of the action, we correct the tree level scaling in Eq. \eqref{eq: rescaling} as follows,
\be
\label{eq: scale transformation 1}
\mathbf{x} \mapsto e^{\delta\ell} \mathbf{x}, 
\qquad
\tau \mapsto e^{z \delta\ell} \tau, 
\qquad 
\vphi \mapsto e^{-\delta\ell/2}Z_\phi^{-1/2}\vphi= e^{-(1/2 + \eta_\phi ) \delta\ell} \vphi,
\ee
where $\eta_\phi$ and $z$ are chosen to cancel the self energy corrections of Eqs.~\eqref{eq: SE int} and~\eqref{eq: SE dis} respectively, 
\be
\eta_\phi =\frac{1}{2}\frac{\delta}{\delta \ell}\log Z_\phi= \frac{2}{3 \pi^2 N}\,,\qquad\quad
z = 1+ \frac{32 \bar \Delta}{\pi N}\,.
\label{eq: z and eta phi}
\ee
Here, $\eta_\phi>0$ is the usual anomalous dimension of $\vphi$ arising from its interaction with $\sigma$ at the clean Wilson-Fisher fixed point \cite{ZinnJustin2002}. 
The deviation of the dynamical exponent $z$ from unity signals the breaking of Lorentz invariance by  quenched disorder. 
In Appendix~\ref{app:zCheck}, we 
check our result for $z$ against a general expression derived in Refs.~\citenum{Aharony2018} and~\citenum{Aharony2018a} for dirty fixed points accessible through conformal perturbation theory.
The agreement between this result and the value of $z$ shown above serves as confirmation of our diagrammatic calculation.

We now 
study the remaining one-loop diagrams, which correct the vertex $\Gamma_{\sigma\phi^\dagger\phi}(\omega=0, |\mathbf{p}|=0)$. 
As shown on the second line of Fig.~\ref{fig: 1/N diagrams}, there are contributions from both interactions and disorder,
\begin{align}
\delta \Gamma_{\sigma \phi^\dagger\phi} &= \delta \Gamma_{\mathrm{int}} + \delta \Gamma_{\mathrm{dis}},\\
\delta \Gamma_{\mathrm{int}} &= \frac{i}{\sqrt{N}} \frac{8}{N} \int^\Lambda_{(1-\delta\ell)\Lambda} \frac{d^2 \mathbf{k}}{(2\pi)^2}\int_{-\infty}^\infty\frac{dk_0}{2\pi} \frac{|k|}{k^4} 
= \frac{i}{\sqrt{N}} \frac{4}{\pi^2 N} \delta\ell\,,\\
\delta \Gamma_{\mathrm{dis}} &= -\frac{i}{\sqrt{N}} \frac{64 \bar \Delta}{N}  \int^\Lambda_{(1-\delta\ell)\Lambda} \frac{d^2 \mathbf{k}}{(2\pi)^2} \frac{|\mathbf{k}|^2}{|\mathbf{k}|^4} = -\frac{i}{\sqrt{N}} \frac{32 \bar\Delta}{\pi N} \delta\ell\,.
\end{align}
Additional $\mathcal{O}(1/N)$ vertex diagrams do exist, but are not logarithmically divergent, as verified in Appendix~\ref{app:DimReg}.
The corrections obtained above must be added to the action $S_r$. 
Imposing scale invariance and the marginality of the $\sigma|\vphi|^2$ vertex requires updating Eq.~\eqref{eq: rescaling} once more to include the anomalous dimension $\eta_\sigma$:
\be
\sigma \mapsto e^{-2\delta\ell}Z_\sigma^{-1/2}\sigma =e^{-(2+\eta_\sigma) \delta\ell} \sigma\,.
\ee
Together with the results for $z$ and $\eta_\phi$ in Eq. \eqref{eq: z and eta phi}
, we find
\be
\eta_\sigma =\frac{1}{2}\frac{\delta\log Z_\sigma}{\delta\ell}= z-1 - 2\eta_\phi + \frac{32\bar\Delta}{\pi N} -\frac{4}{ \pi^2 N}  = \frac{64\bar\Delta}{\pi N} -\frac{16}{3 \pi^2 N}\cdot
\ee
We verify that the second term is the known value of the $\mathcal{O}(1/N)$ anomalous dimension of $\sigma$ at the clean Wilson-Fisher fixed point \cite{Moshe2003}.

\subsubsection{A Dirty Quantum Critical Point}

In light of the comments in Sec.~\ref{sec:ThePlan}, the information obtained in the previous section allows us calculate the running of $\bar{\Delta}$ directly from Eq.~\eqref{eq: disorder action}, which yields
\be
\beta_{\bar \Delta} 
= -\frac{\delta \bar \Delta}{\delta\ell} 
= 2(1-z+\eta_\sigma ) \bar \Delta 
= \left(\frac{64}{\pi} \bar\Delta - \frac{32}{3 \pi^2}  \right) \frac{\bar\Delta}{N}\cdot
\ee
The flows exhibited by this $\b$ function are shown in Fig.~\ref{fig:RGflows}(b).
In particular, a fixed point with both finite disorder and interactions occurs at
\be
\label{eq: fixed point}
\bar\Delta_* = \frac{1}{6 \pi}\cdot
\ee
%
This fixed point constitutes a disordered, interacting quantum critical point! 
It is {\it attractive} (IR stable) in $\bar \Delta$ and ${u}$, but is unstable to perturbations in the mass of the boson, $\delta r\,|\vphi|^2$, which are allowed by symmetry\footnote{We define a quantum critical point as being a fixed point of a RG flow that can be perturbed by relevant operators without explicitly breaking a symmetry. 
This is in contrast to a quantum critical {\it phase}, for which any relevant perturbation breaks a symmetry.}. For $\delta r < 0$, the theory flows to a phase in which the global $\mathrm{O}(2N)$ symmetry is spontaneously broken, and the ground state hosts Goldstone bosons. 
On the other hand, for $\delta r > 0$, the theory flows to an insulating phase. 

The QCP we have obtained is characterized by universal dynamical and correlation length exponents,
\be\label{eqn:zNuResults}
\nu=1\,,\qquad z=1+\frac{16}{3\pi^2N}\,,
\ee
where the correlation length exponent $\nu$ is defined via
\be
\xi\sim|\delta r|^{-\nu}.
\ee
From dimensional analysis, this implies
\be\label{eqn:nuDef}
\nu^{-1}=z+d-[|\vphi|^2]=z-\eta_\sigma=1-\frac{1}{2}\,\beta_{\bar\Delta}\,.
\ee
As we have demonstrated at $\mathcal{O}(1/N)$ [see Appendix~\ref{app:DimReg}], so long as no additional anomalous dimensions are associated with disorder, the $\beta_{\bar{\Delta}}$ is given by Eq.~\eqref{eq: beta function of disorder}, implying that the fixed point condition is identical to the statement $\nu=1$,
\emph{i.e.} $\nu$ receives no quantum corrections.
We can view this as the physical manifestation of the counterbalancing between disorder and interactions at the QCP.
On the other hand, having $1<z<2$ is a reflection of the fact that this is a disordered quantum critical point -- Lorentz invariance is broken, and the density of states $\rho(\varepsilon)\sim\varepsilon^{(d-z)/z}$ vanishes as $\varepsilon\rightarrow0$ at the transition.

Specifying to $N=1$, the symmetry-broken state is a superfluid.
The gapped, symmetry-preserving phase may be the ``Mott glass" phase \cite{Giamarchi2001,Weichman2008}, which is an insulating, glassy state with {\it vanishing} compressibility. This is in contrast to the perhaps more famous Bose glass phase, which includes disorder that does not respect particle-hole ($\PH$) symmetry and has finite compresibility. We comment further on this case in the next subsection, although we emphasize that the glassy nature (or lack thereof) of the disordered state accessible through the dirty QCP derived here cannot be confirmed using our perturbative approach.
Analytic continuation of Eq.~\eqref{eqn:zNuResults} to $N=1$ yields
\be
\label{eq: exponents N=1}
\nu=1\,,\qquad z\approx 1.5\,.
\ee
Remarkably, these results are both consistent with  recent numerical studies of the dirty superfluid-Mott glass transition \cite{Prokofev2004, Vojta2016}. To our knowledge, the quantum critical point we describe here is the only analytic result to achieve this. It is therefore a tantalizing possibility that the fixed point we obtain is in the same universality class as this transition. 




\subsection{Comparison with the Double-$\epsilon$ Expansion}

It is important to 
understand the relationship the dirty QCP examined here has with those obtained in earlier approaches to the dirty boson problem. As mentioned in the Introduction, 
theories of bosons with self-interactions and random mass disorder have been considered before using an expansion in the number of spatial dimensions, $\epsilon=4-d$, and the number of time dimensions, $\epsilon_\tau=d_\tau$ \cite
{Dorogovtsev1980,Cardy1982,Lawrie1984}. This expansion involves perturbing the Gaussian fixed point in $d=4$ dimensions with classical ($d_\tau=0$) disorder, a situation far-removed from the physically relevant case of $d=2$, $d_\tau=1$. 
While this approach also yields a fixed point with finite disorder and interaction strengths, it exhibits some potentially pathological irregularities.
As Fig. \ref{fig:RGflows}(a) demonstrates, upon extrapolating back to $d=2,\,d_\tau=1$, the RG  flows in the critical point's vicinity are spirals for the 
case of a single species of complex bosons ($N=1$). 
In contrast, the results obtained in this paper through a large-$N$ expansion show no indication of spiralling flows. 
This is not necessarily incompatible with the double-$\ep$ expansion since more germane, direct flows similar to Fig.~\ref{fig:RGflows}(b) do appear when $N>N_c=11+6\sqrt{3}\approx21.4$.
Therefore, while we must remain open to the possibility that spiralling flows may appear at a higher order in $1/N$, we argue here that they are instead an artifact of the double-$\ep$ expansion, implying that our results may be more physically relevant even for relatively small values of $N$.

We first note that 
the peculiar flows that appear in the double-$\epsilon$ theory follow from the appearance of complex anomalous dimensions, a signature of non-unitarity
\cite{Aharony2018,Aharony2018a}: 
unlike a unitary theory, the operator dimensions of a disorder-averaged theory are not constrained to the real line\footnote{
For example, replica field theories have central charges which vanish in the replica limit, breaking unitarity, despite the fact that each disorder realization is itself a unitary quantum field theory.}.
Nevertheless, in a perturbative expansion about a unitary theory, operators can only acquire complex scaling dimensions in conjugate pairs,
implying that the (real) scaling dimensions of these operators became identical at some point along the RG flow.
Since the $\phi^4$ operator and the operator associated with the quenched disorder 
have the same scaling dimension at the free, classical fixed point in ($d=4,\,d_\tau=0$) being expanded about in the double-$\ep$ formalism, they can \emph{immediately} mix in such a way that their anomalous dimensions enter the complex plane when disorder is added. 
Conversely, at the large-$N$ fixed point, the scaling of $\abs{\vphi}^2$ and thus the disorder operator is non-perturbatively altered, as indicated by a correlation length exponent $\nu=1$ --- a substantial modification from its free value, $\nu=1/2$.
Our expansion accordingly returns no indication of spiralling flows.

The absence of complex scaling dimensions in our theory may be interpreted as the result of balancing between interactions and disorder at the Wilson-Fisher fixed point.
From this perspective, the ubiquity of strong interactions at the Wilson-Fisher fixed point should always deter (though not completely preclude) the formation of complex scaling dimensions. 
Indeed, the critical exponent $\nu$ differs significantly from its free value even for $N=1$ where $\nu\approx0.67$ \cite{Kos2014}.
It is therefore plausible that the propensity for spiralling flows displayed in the double-$\epsilon$ formalism is an unphysical consequence of starting from a degenerate point and that 
the value of $N_c$ obtained by expanding in $\epsilon$ and $\epsilon_\tau$ is greatly exaggerated compared to the true critical number of species for the onset of spiralling flows.

{The failure of the $\epsilon$ expansion to capture the small-$N$ behavior in such situations 
is not unprecedented.  
The Abelian Higgs model, a theory of complex scalar fields coupled to a fluctuating gauge field,
appears to lack a (real) fixed point for $N\leq182$ in $D=4-\epsilon$ spacetime dimensions \cite{HalperinLubenskyMa1974}. However,  lattice duality with the 3$d$ XY model \cite{Peskin:1977kp,Stone1978,DasguptaHalperin1981}, for which the critical theory is the Wilson-Fisher fixed point discussed here, and numerical results \cite{Nguyen1999,KAJANTIE2004} place that critical number at values as small as one. 
As in the dirty boson problem, this phenomenon can be traced to the presence of two operators having the same scaling dimension.}  

We caution that while the agreement of our results with numerics is indeed remarkable, the arguments outlined by no means constitute a proof that the large-$N$ expansion offers any advantage over the double-$\epsilon$ treatment or even that it is physically relevant.
For $N=1$, both methods are predicated on the disconcerting assignment of a small expansion parameter to an $\mathcal{O}(1)$ value, and both are therefore fundamentally suspect in this regime.
We acknowledge that the absence of spiralling flows and complex dimensions in our study may simply follow from the fact we are perturbing about the regime where the flows from the Wilson-Fisher fixed point are regular.
Nevertheless, even were this the case, our treatment and the fixed point should remain valid at least for sufficiently large-$N$.

\section{Scalar and Vector Potential Disorder}\label{sec:CurrentDisorder}

We have so far focused exclusively on theories that preserve a global $\mathrm{U}(N)$, time-reversal ($\T$), and particle-hole ($\PH$) symmetry for each realization of disorder, and we have shown that this is equivalent to imposing a global $\mathrm{O}(2N)$ symmetry.
In this section, we relax this constraint by only imposing the discrete $\T$ and $\PH$ symmetries on average, allowing for additional disorder perturbations.
Such perturbations can be chosen to preserve the $\mathrm{U}(N)$ symmetry for each disorder realization, but not the $\mathrm{O}(2N)$ symmetry.

The symmetries $\PH$ and $\T$ are broken respectively by random scalar and vector potentials, which we denote $\mathcal{V}(\mathbf{x})$ and $\mathcal{A}_i(\mathbf{x})$, 
\eq{
\mathcal{L}_{J\text{-}\mathrm{dis}}
&=
\V(\vb{x})\,J_\tau(\vb{x},\tau)
+
\sum_{i=x,y}\,\mathcal{A}_i(\vb{x})\,J_i(\vb{x},\tau)
}
where
\eq{
J_\tau
&=
\vphi^\dagger{\pd_\tau}\vphi - \pd_\tau\vphi^\dagger\,\vphi,
&
J_i
&=
i\left(\vphi^\dagger{\pd_i}\vphi - \pd_i\vphi^\dagger\,\vphi\right).
\label{eq:random-scalar-vector-potentials}
}
Here, the scalar potential disorder may be interpreted as a random chemical potential that breaks $\PH$, while vector potential disorder can be associated with a random magnetic flux that breaks $\T$ and parity ($\mathcald{P}$). 
The current $J_\mu$ is the global current corresponding to the electromagnetic charge, a $\mathrm{U}(1)$ subgroup of the global $\mathrm{U}(N)$ symmetry. While it may also be interesting to study disorder that couples to non-Abelian $\mathrm{U}(N)$ currents, such disorder breaks the $\mathrm{U}(N)$ symmetry within each  realization, so we do not consider it.

As for the random mass disorder discussed in the previous section, we assume that scalar and vector potential disorder is drawn from a Gaussian white noise distribution with zero mean,
\begin{align}
\label{eq: scalar/vector disorder correlations}
\overline{\V(\mathbf{x})\V(\mathbf{x}')}&=\Delta_\V\,\delta(\mathbf{x}-\mathbf{x}')\,,
&
\overline{\mathcal{A}_i(\mathbf{x})\mathcal{A}_j(\mathbf{x}')}&=\Delta_\mathcal{A}\,\delta_{ij}\,\delta(\mathbf{x}-\mathbf{x}')\,,
&
\overline{\V(\mathbf{x})}&=\overline{\mathcal{A}_i(\mathbf{x})}=0.
\end{align}
The case of general disorder correlations can also be studied, although we limit ourselves to the Gaussian white noise case for clarity. 

Because $\V$ and $\mathcal{A}_{x,y}$ respectively couple to the temporal and spatial components of a conserved (Abelian) global current, their scaling dimensions satisfy certain non-perturbative constraints, and we use these to derive stability criteria that hold even away from a critical point.
While $[J_\tau]=[J_i]=2$ for relativistic ($z=1$) theories in 2+1 dimensions, these relations are modified in the absence of Lorentz symmetry.
To see how, we recall that the currents' dimensions are fixed by their conservation,
\be
\label{eq: current conservation}
\pd_\mu J^\mu=0\,,
\ee 
which implies a conserved, {\it dimensionless} charge 
\be
\label{eq: charge}
Q=\int d^2\mathbf{x}\,J^\tau\,.
\ee
More precisely, in the quantum theory, current conservation is the statement that correlation functions of $J_\mu$ satisfy Ward identities that embody the condition~\eqref{eq: current conservation}. 
The requirement that $Q$ in Eq.~\eqref{eq: charge} be dimensionless returns
\be
\label{eq: dim of J0}
[J_\tau]=2.
\ee
while the continuity equation, Eq. \eqref{eq: current conservation}, indicates that $\pd_\tau J^\tau$ and $\pd_i J^i$ must have the same scaling dimension, which gives
\be
\label{eq: dim of Ji}
[J_i]=1+z\,.
\ee
Armed with the knowledge that any disorder leads to a deviation of $z$ above unity, we use these relations to deduce the running of $\Delta_\V$ and $\Delta_{\mathcal{A}}$, both near the clean Wilson-Fisher fixed point and the dirty quantum critical point obtained in the previous subsection.

We first consider the case of vector potential disorder in the absence of scalar potential disorder.
From Eq.~\eqref{eq: dim of Ji}, dimensional analysis indicates that $[\mathcal{A}]=1$, which should be familiar as the usual scaling dimension of a vector potential. 
We conclude from Eq.~\eqref{eq: scalar/vector disorder correlations} that $[\Delta_{\mathcal{A}}]=0$ to \emph{all} orders.
Phrased in terms of $\beta$-functions, this reads simply as
\be
\beta_{\Delta_{\mathcal{A}}}=0\,.
\ee 
In other words, the random vector potential is \emph{exactly} marginal, both at the clean Wilson-Fisher fixed point and at our dirty quantum critical point. 
No matter how the dynamical exponent $z$ is renormalized, $\Delta_{\mathcal{A}}$ will not run, resulting in a fixed line parameterized by $z$.  

We now turn to the random scalar potential, following the same logic as we did for vector potential disorder.
Using the fact that Eq.~\eqref{eq: dim of J0} implies $[\V]=z$, together with Eq.~\eqref{eq: scalar/vector disorder correlations}, we find $[\Delta_\V]=2z-2$, which is equivalent to
\be
\beta_{\Delta_\V}
=-(2z-2)\Delta_\V\,.
\ee
Hence, $\Delta_\V$ is {\it relevant} for any $z>1$: both the clean Wilson-Fisher fixed point and our dirty quantum critical point are unstable to $\Delta_\V$, regardless of the strength of the mass or vector potential disorder. 

Although the theory flows to strong disorder, and its ultimate fate cannot be understood perturbatively, one can speculate that the theory flows to a glassy state. Since $\PH$ is broken in each realization, this may be the Bose glass, which has finite compressibility despite being an insulator \cite{Fisher1988,Fisher1989}. Indeed, the exponents we obtain in Eq. \eqref{eq: exponents N=1} are fairly close to those obtained for the disorder-tuned transition between a superfluid and Bose glass if $\PH$ is only imposed on average \cite{Priyadarshee2006, Meier2012, Pollet2009, Ng2015}. In particular, $\nu=1$ is always seen, although there appears to be some disagreement in $z$\footnote{For many years, it was expected that the superfluid-Bose glass transition in $d$ spatial dimensions should have $z=d$ because both phases have finite compressibility, which scales in temperature like $\kappa=\pd n/\pd\mu\sim T^{(d-z)/z}$ \cite{Fisher1988,Fisher1989}. 
However, this expectation relies on the assumption that the measured compressibility is determined by the singular part of the free energy, which is not always the case \cite{Weichman2007}.}. 
This indicates that the quantum critical point obtained in the previous subsection may at least be in a similar universality class to these transitions.


The conclusions of this section hold in general for quenched disorder that couples to conserved Abelian global currents. 
The exact marginality of the random vector potential and the relevance of the random scalar potential for $z>1$ are already well-known in the context of dirty non-interacting Dirac fermion systems \cite{Fradkin1986,Fradkin1986a,Ludwig1994}. 
They were also understood in the strongly interacting context of QED$_3$; there, the global $\mathrm{U}(1)$ current is actually a monopole current,
$j^\mu=\varepsilon^{\mu\nu\lambda}\pd_\nu a_\lambda/2\pi$, where $a$ is the fluctuating gauge field, and so random density and random flux exchange roles \cite{Goswami2017,Thomson2017,Goldman2017}. 
Note that if we had introduced disorder in the non-Abelian $\mathrm{U}(N)$ currents, this would have broken the $\mathrm{U}(N)$ symmetry explicitly in each realization, invalidating the non-perturbative conclusions of this section. 

\section{\texorpdfstring{Boson-Fermion Duality and the $N=1$ Theory}{Duality and the N=1 Theory}}\label{sec:Duality}

The proposal of a web of dualities connecting a menagerie of quantum critical points and phases in 2+1 spacetime dimensions \cite{Seiberg2016, Karch2016} has resulted in progress on several condensed matter problems \cite{Son2015,Wang2015,Metlitski2016,Radicevic2016,Goldman2017,Hui2019,Wang2017a,Thomson2018,Goldman2018a,Goldman2019}. 
These dualities are non-perturbative tools that enable one to determine the low-energy behavior of a strongly-coupled quantum field theory by instead considering the physics of a dual theory that may be more tractable. In this section, we continue the  results of Sections~ \ref{sec:RG} and~\ref{sec:CurrentDisorder} to the case of $N=1$ and explore their implications for the duals of this theory, following the philosophy of Ref.~\citenum{Goldman2017}. In particular, we focus on the particular case of boson-fermion duality \cite{Polyakov1988,Seiberg2016, Karch2016}, in which the dual theory consists of Dirac fermions coupled to an emergent Chern-Simons gauge field. In Appendix \ref{appendix:Boson-Vortex}, we also consider the case of boson-vortex duality \cite{Stone1978,Peskin:1977kp,DasguptaHalperin1981}, in which the dual theory, known as the Abelian Higgs model, consists of bosonic vortices coupled to a fluctuating emergent gauge field. 
In both cases, an immediate consequence of the duality is that, in the presence of a random mass, the dual theory flows to a dirty, interacting QCP with the same exponents as those obtained in Section \ref{sec:RG},
\be
\nu=1\,,\qquad z=1+\frac{16}{3\pi^2}\approx1.5\,.
\ee
We emphasize, however, that this result relies on the extrapolation of $N$ to unity, which may not be valid.


Although many of the results presented in this section are based on conjecture, they nevertheless represent progress in our understanding of dirty Chern-Simons-Dirac fermion theories. 
While disorder has been studied in such theories 
in the limit of a large number of Dirac fermion species \cite{Ye1999}, such expansions suppress the role of the Chern-Simons term to sub-leading orders in $1/N$. The resulting analysis therefore likely misses some of the important global effects of a $\mathcal{O}(1)$ Chern-Simons term. Using duality with the Wilson-Fisher theory circumvents the difficulties of developing a perturbative approach that treats both the disorder and the Chern-Simons gauge field equitably. 

We organize this section as follows. We begin with a brief review of the boson-fermion duality. We next apply the results of Section \ref{sec:RG} for Wilson-Fisher bosons with random mass disorder to the Dirac fermion theory. Finally, we use the non-perturbative results of Section \ref{sec:CurrentDisorder} to comment on the fate of the Dirac theory in the presence of random scalar and vector potentials.


\subsection{Review of the Duality}

We consider the boson-fermion duality \cite{Polyakov1988,Seiberg2016, Karch2016} that relates the Wilson-Fisher theory of the boson $\phi$ to a theory of a Dirac fermion, $\psi$, coupled\footnote{Note that we approximate the Atiyah-Patodi-Singer $\eta$-invariant by a level-$1/2$ Chern-Simons term and include it in the Lagrangian.} to a fluctuating $\mathrm{U}(1)$ Chern-Simons gauge field, $b_\mu$,
\be\label{eqn:WF-fermionDuality}
\mathcal{L}_{\phi}=|D_A\phi|^2-|\phi|^4\longleftrightarrow\mathcal{L}_\psi=i\bar\psi\slashed{D}_b\psi+\frac{1}{8\pi}bdb-\frac{1}{4g^2}f_{\mu\nu}f^{\mu\nu}+\frac{1}{2\pi}bdA+\frac{1}{4\pi}AdA\,,
\ee
The expressions $D_B$, $A\,dB$, $f_{\mu\nu}$, and $\slashed{D}$ are shorthand for $\ptl-iB$, $\varepsilon^{\mu\nu\lambda}A_\mu\ptl_\nu B_\lambda$, and $\ptl_\mu b_\nu-\ptl_\nu b_\mu$, and $D_\mu\gamma^\mu$, respectively.
The double arrow, `$\longleftrightarrow$,' denotes duality.  Since the duality holds only at energy scales much smaller than $g^2$, we omit the Maxwell term, $-\frac{1}{4g^2}f_{\mu\nu}f^{\mu\nu}$, below. For convenience, throughout this section we work with theories in Minkowski spacetime, which are related to the theories considered in earlier sections through a Wick rotation. 
Note that while $\T$ and $\PH$ are manifestly global symmetries of the bosonic theory, $\mathcal{L}_\phi$, they are not immediately apparent in the Dirac fermion theory, $\mathcal{L}_\psi$. 
 Instead, they are to be viewed as emergent IR symmetries of the fermionic theory. Indeed, under this duality, the $\T$ symmetry actually manifests as fermion-vortex self-duality \cite{Seiberg2016}. 

Varying both sides of Eq.~\eqref{eqn:WF-fermionDuality} with respect to $A$, we see that charge in the bosonic theory maps to flux in the fermionic theory,
\be
J^\mu_\phi=i(\phi^\dagger\pd^\mu\phi-\pd^\mu\phi^\dagger\phi)\longleftrightarrow \frac{1}{2\pi}\varepsilon^{\mu\nu\lambda}\pd_\nu\left(b_\lambda+A_\lambda\right)\,,
\label{eq: fermionization-currents}
\ee
where we have introduced the subscript on $J_\phi^\mu$ for clarity.
The physical interpretation of this relation is informed by the flux attachment implemented by the Chern-Simons gauge field. In the fermion theory, charge and flux are slaved to one another through the Chern-Simons gauge field, as are current and electric field. 
Indeed, differentiating the fermion Lagrangrian $\mathcal{L}_\psi$ with respect to $b_\mu$ one finds the mean field equations
\be
\langle\bar\psi\gamma^\mu\psi\rangle+\frac{1}{2}\frac{1}{2\pi}\langle\varepsilon^{\mu\nu\lambda}\pd_\nu b_\lambda\rangle=-\frac{1}{2\pi}\varepsilon^{\mu\nu\lambda}\pd_\nu A_\lambda\,,
\ee
where brackets are used to emphasize that the right-hand side is not an operator, but a c-number. By defining the emergent and background electromagnetic fields $b_*=\varepsilon^{ij}\pd_ib_j, e_i=f_{it}(b), B=\varepsilon^{ij}\pd_iA_j, E_i=\pd_iA_t-\pd_tA_i$, and the Dirac fermion density and current, $\rho_\psi=J_\psi^t=\psi^\dagger\psi, J_\psi^i=\bar\psi\gamma^i\psi$,
we re-express this relation as
\begin{align}
\label{eq: fermion-density-B}
\langle\rho_\psi\rangle+\frac{1}{2}\frac{1}{2\pi}\langle b_*\rangle &= -\frac{1}{2\pi}\,B\,,\\
\langle J_\psi^i\rangle+\frac{1}{2}\frac{1}{2\pi}\varepsilon^{ij}\langle e_j\rangle&=-\frac{1}{2\pi}\varepsilon^{ij}E_j\,.
\label{eq: fermion-current-E}
\end{align} 
The first equation relates the Dirac fermion charge density, $\rho_\psi$, to the sum of the emergent and background magnetic fields, while the second relates the Dirac fermion 
current to the sum of the emergent and background electric fields. 
In contrast, in a typical electromagnetic theory, vector potentials
are associated with currents and scalar potentials are associated with charge.   

It is helpful to determine the relationship between the conductivities of the bosons and fermions, defined via $\langle J^i_\phi\rangle=\sigma^\phi_{ij} E_j$ and $\langle J_\psi^i\rangle=\sigma^\psi_{ij}\langle e^j\rangle$. Combining these definitions with Eqs. \eqref{eq: fermionization-currents} and \eqref{eq: fermion-current-E}, we obtain
\be
\sigma^\psi=-\frac{1}{2}\,\frac{1}{2\pi}\,\varepsilon-\frac{1}{(2\pi)^2}\,\varepsilon\left(\sigma^\phi-\frac{1}{2\pi}\varepsilon\right)^{-1}\varepsilon\,,
\label{eq: Dirac-conductivity-dictionary}
\ee
where tensor indices have been suppressed to reduce clutter. Assuming rotational invariance and expanding in components, this relation becomes
\be
\sigma_{xx}^\psi=\frac{1}{(2\pi)^2}\frac{\sigma_{xx}^\phi}{(\sigma_{xx}^\phi)^2+(\sigma_{xy}^\phi-1/2\pi)^2}\cCom
\quad
\sigma_{xy}^{\psi}=-\frac{1}{2}\frac{1}{2\pi}+\frac{1}{(2\pi)^2}\frac{1/2\pi-\sigma_{xy}^\phi}{(\sigma_{xx}^\phi)^2+(\sigma_{xy}^\phi-1/2\pi)^2}\cdot
\label{eq:Dirac-conductivity-dictionary-componets}
\ee
Since we consider the bosonic theory in the absence of background magnetic fields, we take $\sigma_{xy}^\phi=0$ below.

In terms of the Dirac fermion variables, the superfluid-insulator transition of the bosonic theory is experienced as a  quantum Hall plateau transition tuned by the mass term, $-M\bar{\psi}\psi$.
Integrating out the fermions yields a parity anomaly term for the emergent gauge field, $\sgn(M)\frac{1}{8\pi}b\,db$. 
For $M>0$, the anomaly adds to the Chern-Simons term already in the Lagrangian, which gives the gauge field a so-called `topological mass.' By integrating out the gauge field, 
we see that this state is a trivial, gapped insulator.
To verify that the bosonic dual is also a trivial insulator, we set $\sigma^\phi_{xx}=\sigma^\phi_{xy}=0$ in Eq.~\eqref{eq:Dirac-conductivity-dictionary-componets}, which implies the expected response $\sigma^\psi_{xx}=0$, $\sigma_{xy}^\psi=\nicefrac{+1}{2\cdot2\pi}$.
On the other hand, for $M<0$, the Chern-Simons terms cancel.
The resulting Lagrangian consists of a gapless gauge field $b$, which Higgses the background fields $A$ through the BF term, suggesting that this side of the transition corresponds to the superfluid phase, with $b$ acting as the dual to the Goldstone mode of the bosonic theory.
The insertion of the expected bosonic response, $\sigma_{xx}^\phi\to\infty$, $\sigma^\phi_{xy}=0$, into Eq.~\eqref{eq:Dirac-conductivity-dictionary-componets} accordingly yields $\sigma_{xx}^\psi=0$, $\sigma^\psi_{xy}=\nicefrac{-1}{2\cdot2\pi}$.
We therefore conclude that, as in boson-vortex duality, the mass operators of the two theories are dual to one another,
\be
|\phi|^2\longleftrightarrow\bar\psi\psi\,.
\label{eq: Fermion-mass-dictionary}
\ee  
This operator duality is highly non-trivial: it implies that $\bar\psi\psi$ has the same dimension as $|\phi|^2$ at the Wilson-Fisher fixed point, $[|\phi|^2]\approx 1.5$, meaning that interactions with the Chern-Simons gauge field lead to a {\it negative} anomalous dimension at the clean fixed point, $\eta_{\bar\psi\psi}\sim-0.5$.

\subsection{Random Mass}

Having reviewed the boson-fermion duality in the clean case, we now consider the effects of quenched disorder (again with Gaussian white noise correlations) in the Dirac fermion theory in the absence of the background field $A$. We mention that, since the boson-fermion duality is valid only in the IR, we require the disorder to be sufficiently long-wavelength that it may be considered a perturbation of the IR fixed point.

We first study the effect of a random mass. 
From Eq.~\eqref{eq: Fermion-mass-dictionary}, we again find that mass disorder maps to mass disorder 
\be
R(\mathbf{x})\,|\phi|^2(\mathbf{x},t)\longleftrightarrow R(\mathbf{x})\,\bar\psi\psi(\mathbf{x},t)\,.
\ee
As described in Section \ref{sec:RG}, a random mass causes the bosonic theory in the large-$N$ limit to flow to a disordered, interacting QCP. Provided this remains true for $N=1$, duality implies that the Dirac fermion theory also flows to such a QCP and that at this fixed point, the Dirac fermion mass operator has scaling dimension,
\be
[\bar\psi\psi]=[|\phi|^2]=2+\frac{3}{16\pi^2}\cdot
\ee
Moreover, the identification of the QCPs across the duality also implies that the correlation length and dynamical scaling exponents of the Dirac theory, respectively denoted $\nu_\psi$ and $z_\psi$, are identical to those obtained in Section \ref{sec:RG},
\be
\nu_\psi=\nu=1\,,\qquad z_\psi=z=1+\frac{16}{3\pi^2}\approx1.5\,.
\ee
To our knowledge, no quantum critical point of this type has been obtained perturbatively in Chern-Simons-fermion theories. While the problem of mass disorder in Chern-Simons-Dirac fermion theories was studied in a large-$N$ limit by Ye \cite{Ye1999}, he found that a random mass was marginally irrelevant in the absence of Coulomb interactions. 

Since the QCP studied here is characterized by a universal DC conductivity, it would be very interesting to 
determine the DC transport properties of the Dirac fermions by applying the transport dictionary, Eq. \eqref{eq:Dirac-conductivity-dictionary-componets}, utilizing the DC response of the Wilson-Fisher bosons with a random mass. However, we leave this calculation, which is possible both using a large-$N$ approach and numerical techniques, for future work. 

\subsection{Random Scalar and Vector Potentials}

We now introduce random scalar and vector potentials, as in Eq.~\eqref{eq:random-scalar-vector-potentials}. We emphasize that the conclusions of this section are non-perturbative, and so are valid for $N=1$. They are also consistent with the results of Ye \cite{Ye1999} when Coulomb interactions are turned off. From the current mapping, Eq. \eqref{eq: fermionization-currents}, we first see that a random chemical potential in the bosonic theory maps to a randomly sourced flux in the Dirac fermion theory,
\be
\mathcal{V}(\mathbf{x})\,J_0(\mathbf{x},t)\longleftrightarrow\frac{1}{2\pi}\mathcal{V}(\mathbf{x})\,\varepsilon^{ij}\pd_ib_j(\mathbf{x},t)\,.
\ee
Importantly, the flux attachment constraint, Eq. \eqref{eq: fermion-density-B} implies that randomly sourcing the emergent magnetic field is equivalent to randomly sourcing the Dirac fermion density since the two operators are identical in the absence of an external magnetic field, $B=0$. 
In other words, this disorder should be simultaneously understood as a random current and a random chemical potential (electric field), as can be seen from Eq.~\eqref{eq: fermion-current-E} by noting that a random scalar potential corresponds to $E_j=\pd_j\mathcal{V}/2\pi$.  

From Section \ref{sec:CurrentDisorder}, we recognize that a random scalar potential is relevant, and we expect its addition to push the bosonic theory towards an insulating and possibly glassy phase. If this is true, then the DC response of the bosons is $\sigma_{xx}^\phi=\sigma_{xy}^\phi=0$.
The dual fermions therefore exhibit the same Hall effect as in the clean insulating state, $\sigma_{xx}^\psi=0,\sigma^\psi_{xy}=+\frac{1}{2}\frac{1}{2\pi}$. 
It would be interesting to improve our understanding of this state in future work.

We conclude this section by considering a random vector potential, 
\be
\mathcal{A}^i(\mathbf{x})\,J_i(\mathbf{x},t)\longleftrightarrow\frac{1}{2\pi}\,\mathcal{B}(\mathbf{x})\,a_t(\mathbf{x},t)
\ee  
where $\mathcal{B}=\varepsilon^{ij}\pd_i\mathcal{A}_j$. From Eq. \eqref{eq: fermion-density-B}, the random field $\mathcal{B}(\mathbf{x})$ should be interpreted both as a random density and a random random vector potential (magnetic field). As we observed in Section \ref{sec:CurrentDisorder}, this kind of perturbation is exactly marginal in the bosonic theory, and so the same should hold in the fermionic dual. 

\section{Discussion}
\label{sec: discussion}

In this work, we have revisited the problem of quenched disorder at the quantum superfluid-insulator transition by directly introducing disorder at the strongly coupled Wilson-Fisher fixed point of the $\mathrm{O}(2N)$ model in $2+1$ spacetime dimensions. Using a controlled large-$N$ expansion, we showed that, in the presence of a quenched random mass, the Wilson-Fisher fixed point flows directly to a QCP characterized by finite disorder and interaction strengths. When $N$ is extrapolated to unity, the critical exponents for this transition are strikingly close to recent numerical results for the superfluid-Mott glass transition. 
As far as we are aware,
ours is the first construction to achieve this, indicating that the QCP we obtain may be in the same universality class as the superfluid-Mott glass transition. This is in contrast to earlier approaches using the double-$\epsilon$ expansions about the non-interacting fixed point, which returns spiralling RG flows that are not of obvious physical significance.
Indeed, the relative simplicity of our result is a testament to the important roles played by both strong interactions and disorder in 2$d$ quantum critical systems.

In addition, we presented non-perturbative results for the stability of this QCP to random scalar and vector potentials. While a random vector potential is exactly marginal, a random scalar potential is relevant, leading to what is likely a kind of compressible, glassy state referred to as a Bose glass. Understanding the nature of this glassy state and its relationship to the phenomenology of the Bose glass is an interesting direction for future exploration, although it requires accounting for non-perturbative, rare region effects. 
The theories considered in this work may provide interesting platforms for the study of such non-perturbative effects when both disorder and interactions are present.

By setting $N$ to unity and applying our results 
to dual theories of a Dirac fermion coupled to a fluctuating Chern-Simons gauge field, as well as the Abelian Higgs model (in Appendix \ref{appendix:Boson-Vortex}), we were able to make conjectures regarding the behavior of these theories to quenched disorder. 
Our conclusions constitute significant progress in the study of both of these historically difficult problems.
The results of these approaches can then be compared to our conjecture from duality.

In addition to the critical exponents computed here, the QCP we discuss possesses universal DC and optical conductivities. 
Examining the universal transport properties of this theory via analytic of numerical techniques is important for understanding randomness at the Wilson-Fisher fixed point, as well as its duals.
Such information may shed light  on universal features of both superconductor-insulator transitions (the Abelian Higgs model) and plateau transitions (the Chern-Simons-Dirac theory).

\section*{Acknowledgements}

We thank E. Fradkin, Y.-B. Kim, S. Kivelson, S.-S. Lee, J. Maciejko, M. Mulligan, S. Raghu, S. Ryu, S. Sachdev, B. Spivak, T. Vojta, and S. Whitsitt for discussions. HG is supported by the National Science Foundation (NSF) Graduate Research Fellowship Program under Grant No. DGE-1144245. LN is supported by the Kadanoff Fellowship from University of Chicago. ZB is supported through the Pappalardo Fellowship at MIT. 
AT acknowledges support from the Walter Burke Institute for Theoretical Physics at Caltech and the Caltech Institute for Quantum Information and Matter, an NSF Physics Frontiers Center with support of the Gordon and Betty Moore Foundation through Grant GBMF1250.
This work was performed in part at the Aspen Center for Physics, which is supported by National Science Foundation grant PHY-1607611.
Part of this work was initiated at KITP which is supported by the National Science Foundation under Grant No. NSF PHY-1748958.
\begin{appendix}



\section{RG Calculation using Dimensional Regularization}\label{app:DimReg}
\subsection{Renormalization}



Dimensional regularization is a more natural scheme when considering higher loop diagrams, as needed to calculate the $\sigma$ self energy at $\mathcal{O}(1/N)$.
Our method is as follows. 
The action given in Eq.~\eqref{eq: large N effective action} is the bare action.
For convenience, we reproduce it here:
\eq{
S^B_r&=\sum_n\int d^d\vx\,d\t_B\Bigg[
\vphi^\dag_{B,n}\!\( -{\ptl^2\o\ptl \t_B^2}-{\ptl^2\o\ptl \vx^2}\)\vphi_{B,n}
+{1\o2\cdot8}\,\s_{B,n}\!\(-{\ptl^2\o\ptl\t_B^2}-{\ptl^2\o\ptl\vx^2}\)^{\!-1/2}\!\s_{B,n}
\\
&\quad\quad
+{i\o\sqrt{N}}\s_{B,n} \abs{\vphi_{B,n}}^2 
\Bigg]
+{\bDelta_B\o2}\int d^d\vx\sum_n\int d\t_B \,\s_{B,n}(\vx,\t_B)\sum_m\int d\t'_B\,\s_{B,m}(\vx,\t_B).
\notag
}
Notably, we have added a subscript or superscript `$B$' to the fields, coupling constants, and time coordinate to highlight that these are the bare objects and thus not physical.
The spatial dimension is $d=2-\ep$.
The Feynman rules are the same as those shown in Fig.~\ref{fig: Feynman Rules} and given in Eq.~\eqref{eqn:FeynRules} save that these objects should now include a `$B$' subscript (or superscript).

The physical object is the generating functional $\Gamma$, and the theory is renormalized by ensuring its finiteness at each order in $1/N$.
To guarantee that the time direction is being renormalized correctly, it is useful to rederive the relation between the bare and renormalized vertex functions explicitly.
In doing so, we can suppress both replica and $\mathrm{U}(N)$ vector indices since we assume that neither symmetry is broken.
The generating functional is a function of the bare field configuration $\overline{\vphi}_B$ and $\overline{\s}_B$:
\eq{
\Gamma[\bar{\vphi}_{B},\bar{\s}_{B}]&=\sum_{\N,\M=0}^\infty{1\o\N!\M!}\int \prod_{i=1}^{\N+\M}\( d^d\vx_i\,d\t_i^B\)
\Gamma^{(\N,\M)}_B\(\{\vx_i,\t_i^B\}\)
\nt
&\quad\quad
\times\prod_{j=1}^{\N}\bar{\vphi}_B\(\vx_j,\t_j^B\)\cdot\prod_{k=\N+1}^{\N+\M}\bar{\s}_B\(\vx_k,\t_k^B\),
}
where $\Gamma^{(0,0)}=0$ and the $\Delta_B$ dependence is left implicit.
To make contact with the notation of the main text, we note that the vertices $\Gamma_{\s\phi^\dag\phi}=\Gamma^{(1,2)}$, $\Gamma^{(2,0)}=-(G^\phi)^{-1}$ and $\Gamma^{(0,2)}=-(G^\s)^{-1}$. 

As emphasized, the vertex functions $\Gamma_B$ are not finite in the limit that the UV cutoff $\Lambda\to\infty$.
We define the renormalized fields and time as
\eq{\label{eqn:RenormFieldsParam}
\vphi(\vx,\t)&=Z_\phi^{1/2} \vphi_B\(\vx,\t^B\),
&
\s(\vx,\t)&=Z_\s^{1/2}\s_B\(\vx,\t^B\),
&
\t^B&=Z_\t\t,
&
\bDelta_B&=Z_{\bDelta} \bDelta.
}
The renormalization constants can be written as $Z_x=1+\d_x$, $x=\phi,\s,\t,\Delta$, where $\d_x$ is $\O(1/N)$, allowing for a perturbative treatment.
Inserting the renormalized fields into the functional returns
\eq{
\Gamma[\bar{\vphi}_B,\bar{\s}_B]&=\sum_{\N,\M=0}^\infty{1\o \N!\M!}\int \prod_{\ell=1}^{\N+\M}\( d^d\vx_i\,d\t_i\)
Z_\phi^{\N/2}Z_\s^{\M/2}Z^{\N+\M}_\t\Gamma^{(\N,\M)}_B\(\{\vx_\ell,\t_\ell^B\}\)
\nt
&\quad
\times\prod_{n=1}^\N\bar{\phi}\(\vx_n,\t_n\)\cdot\prod_{m=\N+1}^{\N+\M}\bar{\s}\(\vx_{m},\t_{m}\).
}
The renormalized vertex functions are obtained by differentiating $\Gamma$ with respect to $\bar{\phi}$ and $\bar{\s}$.
It follows that
\eq{
\Gamma_R^{(\N,\M)}\[\{\vx_i,\t_i\}\]&=Z_\phi^{\N/2}Z_\s^{\M/2}Z_\t^{\N+\M}\Gamma_B^{(\N,\M)}\[\left\{\vx_i,\t_i^B\right\}\].
}
Finally, since perturbation theory is more efficiently done in momentum space, we Fourier transform to obtain
\eq{
(2\pi)^{d+1}\d^d\big(\textstyle{\sum_i}& \vb{p}_i\big)\d\big(\textstyle{\sum_i} p_{0,i}\big)\Gamma^{(\N,\M)}_R\[\{\vp_i,p_{0,i}\}\]
\nt
&=(2\pi)^{d+1}\d^d\big(\textstyle{\sum_\ell} p_\ell\big)\d\big(\textstyle{\sum_\ell} p_{0,\ell}^B\big)Z_\phi^{\N/2}Z_\s^{\M/2}\Gamma^{(\N,\M)}_B\[\left\{\vp_\ell,p^B_{0,\ell}\right\}\]
\nt
&=(2\pi)^{d+1}\d^d\big(\textstyle{\sum_\ell} p_\ell\big)\d\big(\textstyle{\sum_\ell} p_{0,\ell}\big)Z_\phi^{\N/2}Z_\s^{\M/2}Z_\t\Gamma^{(\N,\M)}_B\[\left\{\vp_\ell,p^B_{0,\ell}\right\}\],
}
where in the second line we used $p_{0,B}=p_0/Z_\t$.
Cancelling the $\d$-functions, we are left with
\eq{
\Gamma^{(\N,\M)}_R\[\{\vp_\ell,p_{0,\ell}\}\]=Z_\phi^{\N/2}Z_\s^{\M/2}Z_\t\Gamma^{(\N,\M)}_B\[\left\{\vp_\ell,p^B_{0,\ell}\right\}\].
}
The renormalization constants are determined by first calculating the bare vertex functions and cancelling all divergences in $\Gamma_B$ with the counterterms $Z_x$. 
Since we use a dimensional regularization scheme ($D=3-\ep$), this is done by defining the $Z$'s such that all $1/\ep$ poles cancel.
(We express these $1/\ep$ poles in terms of the cutoff $\Lambda$ and renormalization scale $\mu$ in Appendix~\ref{app:ScalingFuns}.)

We emphasize that the bare vertex functions must be computed entirely using the bare propagators and vertex functions, as well as time (frequency).
If this is not done, there is a risk of double counting some of the divergences, as we believe was done in Ref.~\citenum{Kim1994}.

At $\mathcal{O}(1/N)$, only three vertex functions, $\Gamma^{(2,0)}_B$, $\Gamma^{(2,1)}_B$, and $\Gamma^{(0,2)}_B$, need be considered. 
We compute these below.
In what follows, all non-log-divergent contributions (\emph{e.g.} all divergences that do not contribute a $1/\ep$ pole) are ignored.

\subsection{Diagrams}

\subsubsection{\texorpdfstring{$\Gamma^{(2,0)}_R$:  $\phi\phi$ self-energy}{Gamma(2,1):  phiphi self-energy}}
\begin{figure}
	\centering
	\includegraphics[width=0.8\textwidth,page=1]{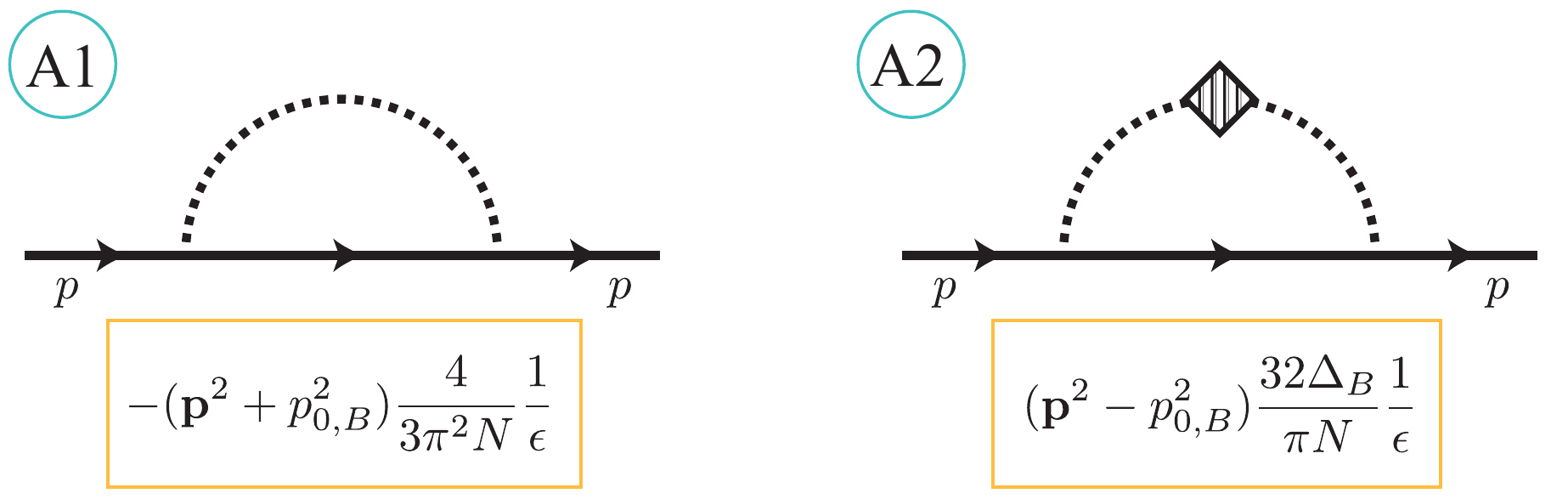}
	\caption{Divergences corresponding to the $\phi$ self energy. 
	In the notation of this appendeix, they contribute to $\Gamma^{(2,0)}_B$.
	The time-component of momenta, $p$, $q$ is considered to be bare, $q=(\vq,q_{0,B})$, etc.
		}
	\label{fig:PhiSelfEnergy}
\end{figure}

The log-divergent contributions to the $\phi$ propagator are shown in Fig.~\ref{fig:PhiSelfEnergy}.
Summing them, we find
\eq{
\Sigma_\phi&=\text{A1}+\text{A2}+\text{finite}
=
-\v{p}^2 \left( \frac{4}{3\pi^2 N}{1\o\ep} - {32\bDelta_B\o \pi N}{1\o\ep} \right)
-p_{0,B}^2 \left( \frac{4}{3\pi^2 N}{1\o\ep} + {32\bDelta_B\o \pi N}{1\o\ep} \right),
}
and then using $\Gamma_B^{(2,0)}=\vp^2+p_{0,B}^2 -\Sigma_{\phi,B}$, gives
\eq{
\Gamma_R^{(2,0)}&=Z_\phi Z_\t\[ \vp^2 \left( 1+ \frac{4}{3\pi^2 N}{1\o\ep} - {32\bDelta\o \pi N}{1\o\ep} \right)
+p_{0,B}^2 \left(1 +\frac{4}{3\pi^2 N}{1\o\ep} + {32\bDelta\o \pi N}{1\o\ep} \right)\]
\nt
&=\vp^2 \left( 1+ \d_\phi+\d_\t+\frac{4}{3\pi^2 N}{1\o\ep} - {32\bDelta\o \pi N}{1\o\ep} \right)
+p_{0}^2 \left(1+ \d_\phi-\d_\t+\frac{4}{3\pi^2 N}{1\o\ep} + {32\bDelta\o \pi N}{1\o\ep} \right).
}
From this we conclude
\eq{\label{eqn:PhiTauCT}
\d_\phi&=-{4\o3\pi^2 N}{1\o\ep}\cCom
&
\d_\t&={32\bDelta\o\pi N}{1\o\ep}\cdot
}

\subsubsection{\texorpdfstring{$\Gamma^{(2,1)}_R$: 3-point vertex}{Gamma(2,1): 3-point vertex}}
\begin{figure}
	\centering
	\includegraphics[width=0.9\textwidth,page=3]{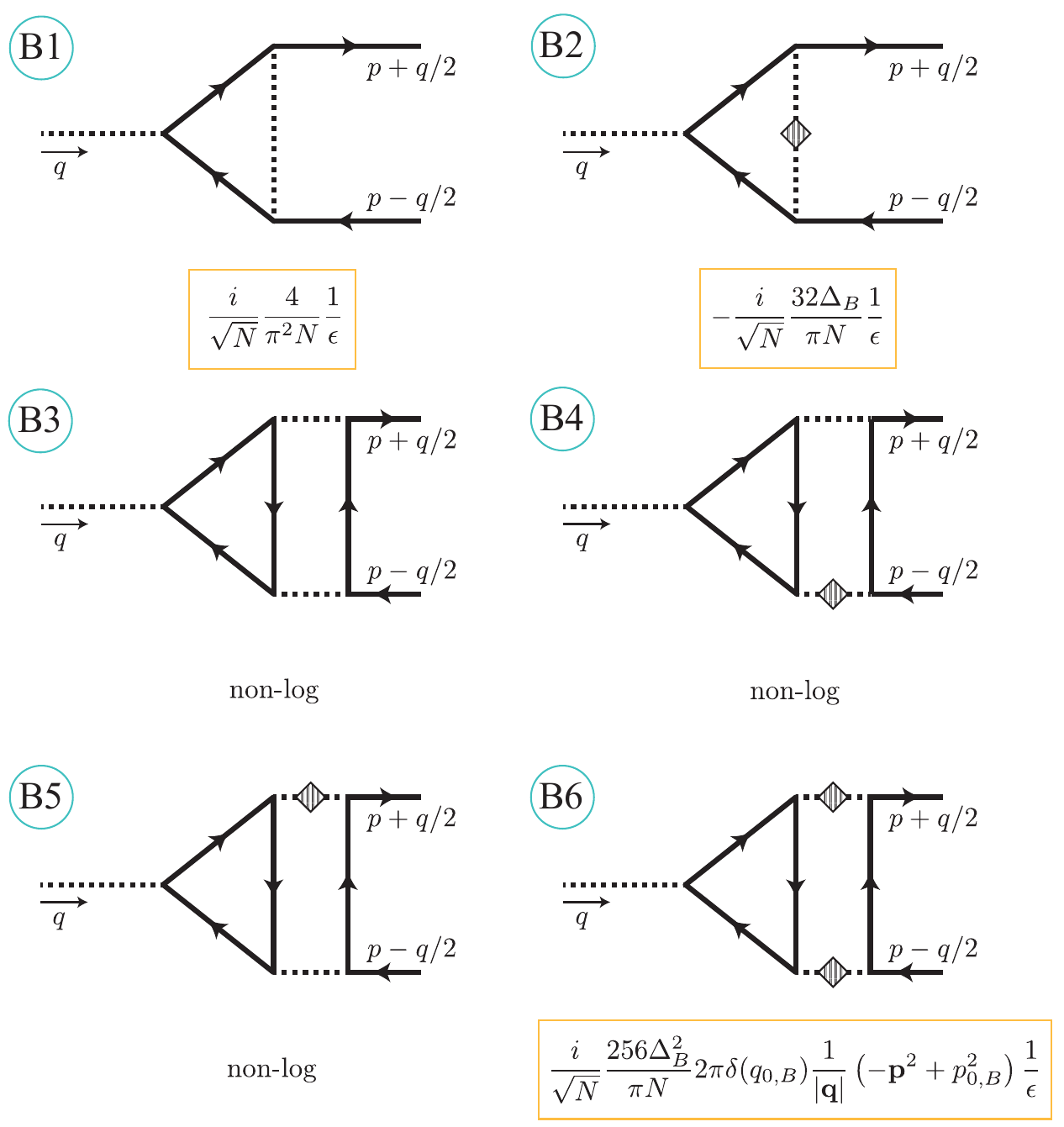}
	\caption{
	All three-point diagrams correcting the $\sigma\abs{\v{\phi}}^2$ vertex, $\Gamma^{(2,1)}_B$ at $\mathcal{O}(1/N)$.
	Diagrams B3-B6 possess partners where the $\phi$ fields traverse the loop in the converse direction.
	The time-component of momenta, $p$, $q$ is considered to be bare, $q=(\vq,q_{0,B})$, etc.
	}
	\label{fig:3ptVertex}
\end{figure}

We summarize the divergent contributions to the 3-point vertex in Fig.~\ref{fig:3ptVertex}. 
We note that the diagram B6 indicates that $\overline{\Braket{\s|\vphi|^2}}$ mixes with 
\eq{
\overline{
\Braket{
\int_\t \, \(-\v{\nabla}^2\)^{-1/2}\s\int_{\t'} \,
\Big[
\Big(\vphi^\dag\v{\nabla}^2\vphi-\v{\nabla}\vphi^\dag\cdot\v{\nabla}\vphi\Big)
-
\Big(\vphi^\dag\ptl_0\vphi-\ptl_0\vphi^\dag\ptl_0\vphi\Big)
\Big]
}},
}
This is a consequence of the fact that the disordered theory is nonrenormalizable.
For the purpose of determining the renormalization constant $Z_\s$, it is not necessary to consider this mixing.

Ignoring these terms, we find that the bare 3-point is
\eq{
\Gamma^{(2,1)}_B&\sim-{i\o\sqrt{N}}+\text{B1}+\text{B2}
=-{i\o\sqrt{N}}
\( 1-{4\o \pi^2 N}{1\o\ep} +{32 \bDelta_B\o\pi N}{1\o\ep}\),
}
implying that the renormalized vertex function is
\eq{
\Gamma_R^{(2,1)}&=Z_\s^{1/2}Z_\phi Z_\t \Gamma_B^{(2,1)}
\nt
&\sim-{i\o\sqrt{N}}
\( 1 +{1\o2}\d_\s  + \d_\phi +\d_\t -{4\o \pi^2 N}{1\o\ep} +{32 \bDelta\o\pi N}{1\o\ep} \)
\nt
&=-{i\o\sqrt{N}}\( 1 +{1\o2}\d_\s -{4\o3\pi^2 N}{1\o\ep} +{32\bDelta\o\pi N}{1\o\ep} -{4\o \pi^2 N}{1\o\ep} +{32 \bDelta\o\pi N}{1\o\ep} \),
}
where the results of Eq.~\eqref{eqn:PhiTauCT} have been inserted in the third line. 
Enforcing the finiteness of $\Gamma_R^{(2,1)}$ requires
\eq{\label{eqn:SigmaCT-3ptVert}
\d_\s&=\({32\o 3\pi^2 N} - {128 \bDelta\o\pi N}\){1\o\ep}\cdot
}

\subsubsection{\texorpdfstring{$\Gamma_R^{(0,2)}$: $\s$ self energy}{Gamma(0,2): sigma self energy}}
\begin{figure}
	\centering
	\includegraphics[width=.85\textwidth,page=4]{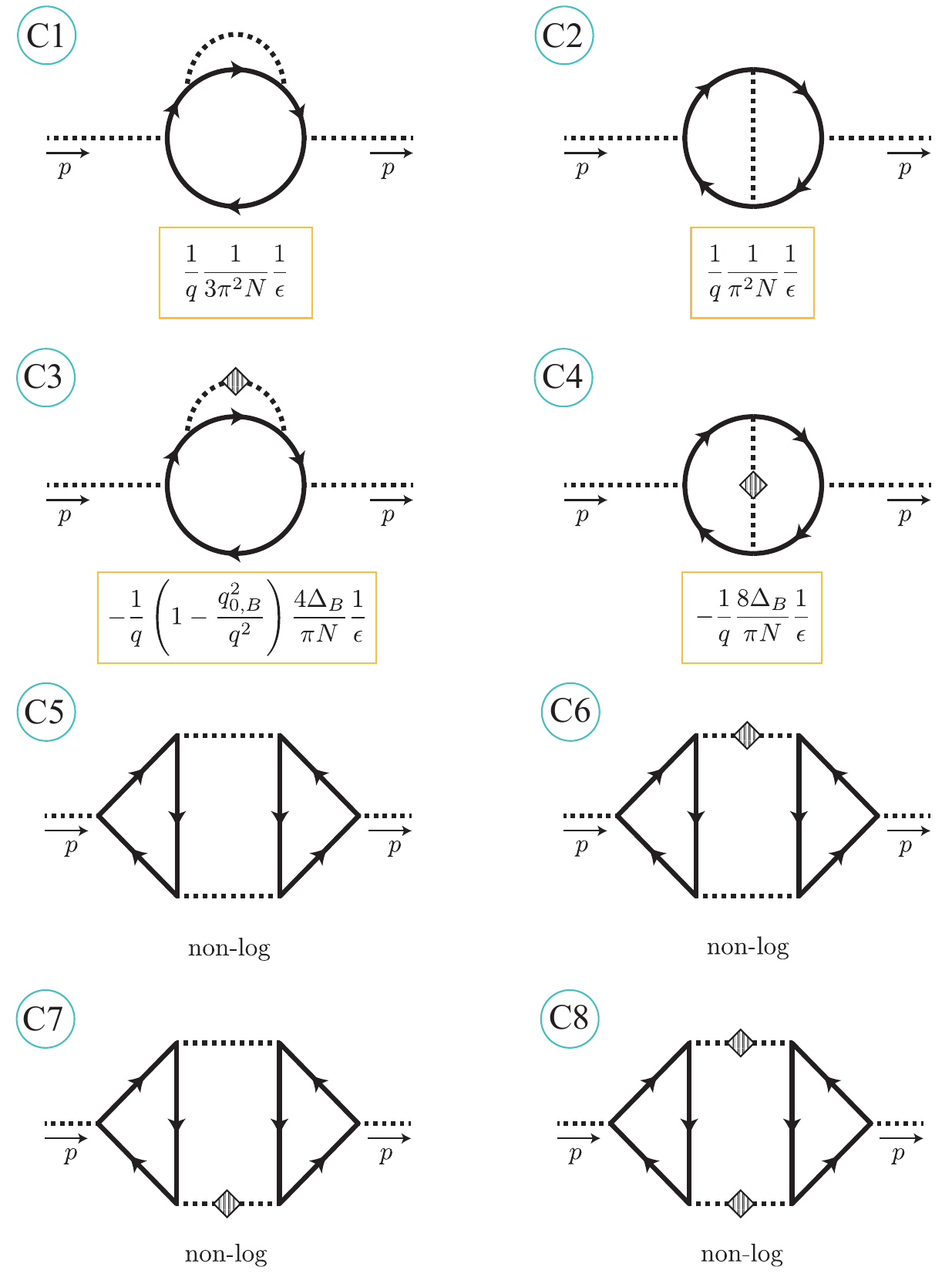}
	\caption{{
	Contributions to the bare $\sigma$ self energy, $\Gamma^{(0,2)}_B=\Gamma_{\sigma\sigma}$ at $\mathcal{O}(1/N)$. Only the two-loop diagrams, C1-C4, are found to contain log divergences. 
	We note that each of these diagrams has a partner where the internal $G^\phi$ lines in the converse direction. These diagrams have been included in the divergent terms shown beneath each diagram.
	The time-component of the momenta, $p$, $q$ is considered to be bare, $q=(\vq,q_{0,B})$, etc.}}
	\label{fig:SigmaSelfEnergy}
\end{figure}

In order to determine whether $\int d\t\,\s(\vx,\t)$ is renormalized differently than $\s(\vx,\t)$, we directly calculate the $\s$ self energy. 
We remark that the renormalization scheme [Eq.~\eqref{eqn:RenormFieldsParam}] cannot account for these types of divergences -- new counterterms would be required. 
Our ability to renormalize $\Gamma^{(0,2)}_R$ with the current set of counterterms is proof that our scheme is sufficient at $\mathcal{O}(1/N)$.
It also serves as verification of our results for $\d_\s$ and $\d_\t$ above.

The log-divergent contributions are shown in Fig.~\ref{fig:SigmaSelfEnergy}.
Adding them, we find
\eq{
\Sigma_{\s,B}[\vq,q_{0,B}]&=\text{C1}+\text{C2}+\text{C3}+\text{C4}
\nt
&={1\o\sqrt{\vq^2+q_{0,B}^2}}\left( {1\o \pi^2 N} +{1\o 3\pi^2 N} - {8\bDelta_B\o\pi N}-{4\bDelta_B\o \pi N} \){1\o\ep}
+{q_{0,B}^2 \o \big(\vq^2 +q_{0,B}^2\big)^{3/2}} {4\bDelta_B\o \pi N}{1\o \ep}
\cdot
}
The bare 2-point $\s$ vertex is therefore
\eq{
\Gamma_B^{(0,2)}[\vq,q_{0,B}]&={1\o 8\sqrt{\vq^2+q_{0,B}^2}}\( 1-{ 32\o 3\pi^2N}{1\o\ep}+{96\bDelta_B\o\pi N}{1\o\ep} - {q_{0,B}^2\o \vq^2 +q_{0,B}^2 }{32\bDelta_B\o \pi N}{1\o\ep}\)
+2\pi\d(q_{0,B})\bDelta_B.
}
To renormalize, we write
\eq{\label{eqn:GammaR02}
\Gamma_R^{(0,2)}[\vq,q_{0}]&=Z_\s Z_\t\Gamma^{(0,2)}_B[\vq,q_{0,B}]
\nt
&={1\o 8\sqrt{\vq^2+q_{0}^2}}
\( 1+\d_\s+\d_\t-{ 32\o 3\pi^2N}{1\o\ep} +{96\bDelta\o\pi N}{1\o\ep} + {q_{0}^2\o \vq^2 +q_{0}^2 }\[\d_\t-{32\bDelta\o \pi N}{1\o\ep}\]\)
\nt
&\quad
+2\pi\d(q_0)\bDelta(1+\d_\s+2\d_\t+\d_\bDelta).
}
Ensuring finiteness returns
\eq{\label{eqn:SigmaTauDeltaCT-SigSelfE}
\d_\s&=\({32\o 3\pi^2 N}-{128\bDelta\o\pi N}\){1\o\ep}\cCom
&
\d_\t&={32\bDelta\o\pi N}{1\o\ep}\cCom
&
\d_\bDelta&=\(-{32\o3\pi^2N}+{64\bDelta\o\pi N}\){1\o\ep}\cdot
}
Our results for $\d_\s$ and $\d_\t$ are notably in agreement with what we obtained from the $\phi$ self-energy and the three-point vertex in Eqs.~\eqref{eqn:PhiTauCT} and~\eqref{eqn:SigmaCT-3ptVert}.

In Ref.~\citenum{Kim1994}, the renormalized $\phi$ propagator was instead used to compute the diagram C3.
As a result, its divergence cancels out and does not appear in Eq.~\eqref{eqn:GammaR02}.

%

\subsection{Scaling functions}\label{app:ScalingFuns}

Summarizing our results from Eqs.~\eqref{eqn:PhiTauCT},~\eqref{eqn:SigmaCT-3ptVert}, and~\eqref{eqn:SigmaTauDeltaCT-SigSelfE}, we have
\eq{
\d_\phi&=-{4\o3\pi^2 N}\[ {1\o\ep} + \log \(\Lambda\o\mu\) \],
&
\d_\tau&={32\bDelta\o \pi N}\[ {1\o\ep} + \log \(\Lambda\o\mu\) \],
\\
\d_\s&=\({32\o3\pi^2 N}-{128\bDelta\o\pi N}\)\[ {1\o\ep} + \log \(\Lambda\o\mu\) \],
&
\d_\bDelta&=\( -{32\o 3\pi^2 N}+{64\bDelta\o \pi N}\)\[ {1\o\ep} + \log \(\Lambda\o\mu\) \].\notag
}
Here, we have taken $\nicefrac{1}{\ep}\to\nicefrac{1}{\ep}+\log\(\nicefrac{\Lambda}{\mu}\)$ through the following reasoning. 
In the Feynamn diagrams calculated, factor of $1/\ep$ is always accompanied by $-\log p$. 
Since $p$ is dimensionful, the logarithm should actually be a fraction of $p$ to some other scale. 
The only other scale in the theory is the UV cutoff $\Lambda$, and it follows that these diagrams should be interpreted as $\#\[ \nicefrac{1}{\ep}+\log\(\nicefrac{\Lambda}{ p}\)\]$ where `\#' represents the coefficients we just calculated.
Hence, in order to ensure that the renormalized diagram is finite as $\Lambda\to\infty$, the $\nicefrac{1}{\ep}$ of the counterterm should be accompanied by $\log\(\nicefrac{\Lambda}{\mu}\)$, where $\mu$ is the renormalization scale: $\delta_j = -\#\[\nicefrac{1}{\ep} +\log\(\nicefrac{\Lambda}{\mu}\)\]$.

With these counterterms, we can now calculate the primary quantities of interest: the dynamical critical exponent $z$, the anomalous dimension for $\phi$, the anomalous dimension for $\s$, and the $\b$-function for the disorder strength $\Delta$. 

\subsubsection{\texorpdfstring{Dynamical critical exponent $z$}{Dynamical critical exponent z
}}
The dynamical critical exponent is defined through
\eq{
\mu {d\o d\mu} \t &=z \t
}
The bare time, conversely, scales as
\eq{
\mu{d\o d\mu}\t_B&=\t_B.
}
Inserting $\tau_B=Z_\tau \tau$, we find
\eq{\label{eqn:DynExp-app}
z&=1-\mu{d\o d\mu} Z_\t = 1+{32\bDelta\o \pi N}\cdot
}

\subsubsection{\texorpdfstring{Anomalous dimensions of $\phi$ and $\s$}{Anomalous dimensions of phi and sigma}}
We define the anomalous dimension of an operator $\O$ as $\eta_\O$ such that $[\O]=[\O]_0+\eta_\O$, where $[\O]_0$ is the engineering dimension of $\O$.
It follows from the definitions of Eq.~\eqref{eqn:RenormFieldsParam} that the anomalous dimension of $\phi$ is
\eq{
\eta_\phi&={1\o2}\mu {d\o d\mu} \log Z_\phi = {2\o 3\pi^2 N}\cCom
}
and of $\s$ is
\eq{
\eta_\s&={1\o2}\mu {d\o d\mu} \log Z_\s = -{16\o3\pi^2 N} + {64\bDelta\o \pi N} \cdot
}
Recall that the operator identity of Eq.~\eqref{eq:OperatorIdentity} implies $\eta_\sigma=\eta_{\abs{\phi}^2}$.

\subsubsection{\texorpdfstring{$\b$ function of $\bDelta$}{Beta function of Delta}}
Finally, $\b_\bDelta$ is defined through the requirement that the bare coupling constant be invariant under RG:
\eq{
\mu {d\o d\mu} \Delta_B&=0.
}
From this we find
\eq{
\b_\bDelta&=\mu{d\o d\mu}\Delta 
=-\bDelta {d\log Z_\bDelta\o d\log \mu}
=- {32\bDelta\o3\pi^2 N}+{64\bDelta^2\o\pi N} .
}
We note that here we are using the high energy convention, so that $\b_\bDelta<0$ implies a flow to strong coupling.


\section{
Check of dynamical critical exponent
}\label{app:zCheck}

The authors of Ref.~\citenum{Aharony2018} derive a formula for the leading order correction to the dynamical critical exponent $z$ of a 
a generic theory with (quantum) disorder of strength $\Delta$ coupling to an operator $\mathcal{O}$.
In Eq.~(4.36) of their paper, they state
\eq{
z-1={\Delta\o2}{c_\mathcal{OO}\o c_T} {D(D+1)\o D-1}{\Gamma(D/2)\o 2\pi^{D/2}}\stackrel{D\to 3}{=}{\Delta\o2}{c_\mathcal{OO}\o c_T}{3\o2\pi}\cdot
}
Here, $c_\mathcal{OO}$ is the coefficient of the two-point $\mathcal{O}$ correlator and $c_T$ is the central charge (the coefficient of the two-point correlator of the stress energy tensor).
We show that this is consistent with our results.

From Eq.~\eqref{eqn:RepActionR}, we see that the disorder couples to $i\sigma(\vx,\t)/u$, and it follows that for us $c_{\mathcal{OO}}=-c_{\s\s}/u^2.$
This coefficient is determined by the real space $\s$ Green's function:
\eq{
G_\s(r)&=\int{d^{D}p\o(2\pi)^{D}} e^{ip\cdot r}\,8\abs{p} = -{8\o\pi^2}{1\o r^4}\cCom
}
implying that
\eq{
-{c_\mathcal{\s\s}\o u^2}={1\o u^2}{8\o\pi^2}\cdot
}
The leading contribution to the central charge of the $\mathrm{O}(2N)$ Wilson-Fisher fixed point corresponds simply to the central charge of $2N$ real, free bosons, which is given by \cite{Petkou95,Petkou96}
\eq{
c_T&\cong 2N\({1\o 2\pi^{D/2}/\Gamma(D/2)}\)^2 {D\o D-1}\stackrel{D\to 3}{=}{3 N\o 16 \pi^2}\cCom
}
where $D=d+1$ is the total number of spacetime dimensions.
Putting this together, we find
\eq{
z-1={1\o2}{\Delta\o u^2}{8/\pi^2\o 3N/16\pi^2}{3\o2\pi}={32\bar{\Delta}\o\pi N}\cCom
}
in perfect agreement with Eq.~\eqref{eq: z and eta phi} (as well as Eq.~\eqref{eqn:DynExp-app} in Appendix~\ref{app:DimReg}).

\section{Boson-Vortex Duality}
\label{appendix:Boson-Vortex}

\subsection{Review of the Duality}

The first duality we consider \cite{Stone1978,Peskin:1977kp,DasguptaHalperin1981} relates a single complex scalar field, $\phi$ (we drop the boldface since $N=1$), at its Wilson-Fisher fixed point to the Abelian Higgs model, a theory of complex bosonic vortices, $\tilde{\phi}$, also at their Wilson-Fisher fixed point. These vortices additionally interact through a logarithmic potential mediated by an emergent $\mathrm{U}(1)$ gauge field, $a_\mu$,
\be
\mathcal{L}_{\phi}=|D_A\phi|^2-|\phi|^4\longleftrightarrow\mathcal{L}_{\tilde\phi}=|D_a\tilde{\phi}|^2-|\tilde{\phi}|^4+\frac{1}{2\pi}Ada-\frac{1}{4g^2}f_{\mu\nu}f^{\mu\nu},
\label{eq: boson-vortex}
\ee
where $A_\mu$ is a background gauge field. 
Here the interaction terms $-\abs{\phi}^4$, $-|{\tilde{\phi}}|^4$ imply that the theories are tuned to the Wilson-Fished fixed point.
As in the case of the boson-fermion duality, we only consider physics at energy scales much smaller than $g^2$, allowing us to omit the Maxwell term, $-\frac{1}{4g^2}f_{\mu\nu}f^{\mu\nu}$. We again work in Minkowski spacetime. 

By differentiating each theory in Eq. \eqref{eq: boson-vortex} with respect to $A_\mu$, one sees that this duality relates charge in the Wilson-Fisher theory to flux in the Abelian Higgs model,
\be
J^\mu=i(\phi^\dagger\pd^\mu\phi-\pd^\mu\,\phi^\dagger\phi)\longleftrightarrow j^\mu=\frac{1}{2\pi}\varepsilon^{\mu\nu\lambda}\pd_\nu a_\lambda\,,
\label{eq:charge-flux1}
\ee
By considering the equations of motion for $a_\mu$ in the Abelian Higgs model, it follows that the converse is also true,
\be
\frac{1}{2\pi}\varepsilon^{\mu\nu\lambda}\pd_\nu A_\lambda=\langle \tilde{J}^\mu\rangle=\langle i(\tilde\phi^\dagger\pd^\mu\tilde\phi-\pd^\mu\,\tilde\phi^\dagger\tilde\phi)\rangle\,.
\label{eq:charge-flux2}
\ee
In terms of global symmetries, the mapping of charge to flux across the duality implies an exchange of $\T$ and $\PH$ symmetries (here defined with appropriate transformation laws for the gauge fields).  
Since current and voltage exchange roles across the duality, the conductivity of the particles $\phi$ corresponds to the resistivity of the vortices $\tilde\phi$ and vice versa
\be
\sigma_{ij}^\phi=\frac{1}{(2\pi)^2}\varepsilon^{ik}\varepsilon^{jl}\rho_{kl}^{\tilde{\phi}}\,,
\label{eq: conductivity dictionary bosons}
\ee
where we write conductivity (resistivity) in units of $e^2/\hbar$ ($\hbar/e^2$). 
This dictionary is obtained using the charge-flux relations, Eqs. \eqref{eq:charge-flux1}-\eqref{eq:charge-flux2}, and the definition of the conductivities $\langle J_i\rangle=\sigma^\phi_{ij}E^j,\langle\tilde{J}_i\rangle=\sigma^{\tilde\phi}_{ij}\langle e^j\rangle$, where $E_i=\pd_iA_t-\pd_tA_i$ and $e(a)=f_{it}(a)$ are the electric fields associated with $A$ and $a$ respectively, and $\rho=\sigma^{-1}$.

The duality, Eq. \eqref{eq: boson-vortex}, can be verified by considering the phase diagrams of each of the dual theories. 
As discussed earlier, the Wilson-Fisher theory is tuned through the addition of a mass, $\delta r\abs{\phi}^2$.
For $\delta r>0$, $\phi$ is gapped, and the ground state is insulating, while for $\delta r<0$, $\phi$ condenses, and the ground state hosts a Goldstone mode. 
On the other hand, when a mass term $-\delta\tilde{r}|\tilde\phi|^2$ with $\d\tilde{r}>0$ is added to the dual theory, $\mathcal{L}_{\tilde\phi}$, $\tilde{\phi}$ is gapped out, but the ground state contains a gapless gauge field. 
This is the superfluid phase seen in in the Wilson-Fisher theory: the gauge field is the dual of the Goldstone mode. Similarly, for $\delta\tilde{r}<0$, $\tilde{\phi}$ condenses and the gauge field is Higgsed, forming a superconductor. 
The conductivity dictionary of Eq. \eqref{eq: conductivity dictionary bosons} indicates that a superconductor of vortices ($\rho^{\tilde\phi}=0$) is an insulator of $\phi$ particles, making it the dual of the insulating phase of $\phi$'s. This mapping of the phase diagrams suggests that the mass operators in the two theories are dual to one another up to a sign,
\be
|\phi|^2\longleftrightarrow-|\tilde\phi|^2\,.
\label{eq:mass dictionary}
\ee
In summary, when the charge in one theory is gapped, the vortices of the dual theory condense, and vice versa.

\subsection{Random Mass}

We now use the results of Sections \ref{sec:RG} and \ref{sec:CurrentDisorder} and the operator dictionaries, Eqs. \eqref{eq:charge-flux1} and \eqref{eq:mass dictionary}, to determine the effects of disorder on the Abelian Higgs model (setting the background field, $A_\mu$, to zero). We begin by considering the effect of a random mass with Gaussian white noise correlations, as discussed in Section \ref{sec:RG}. From Eq. \eqref{eq:mass dictionary}, we see that a random mass at the $N=1$ Wilson-Fisher fixed point is dual to a random mass in the Abelian Higgs model,
\be
R(\mathbf{x})|\phi|^2(\mathbf{x},\tau)\longleftrightarrow-R(\mathbf{x})|\tilde{\phi}|^2(\mathbf{x},\tau)\,.
\ee
Since $R$ is a random variable which can take positive and negative values, the change in sign is immaterial. In the large-$N$ limit, we observed that the Wilson-Fisher fixed point gives way to a QCP with finite disorder and interaction strengths. Under the assumption that this story continues to hold down to $N=1$, the Abelian Higgs model with a random mass must also flow to such a QCP. Moreover, since the mass operators in the two theories are dual to one another, they have the same scaling dimension at the fixed point,
\be
[|\tilde{\phi}|^2]=[|\phi|^2]=2+\frac{3}{16\pi^2}\,.
\ee
As in the boson-fermion duality, the dynamical scaling exponent, $\tilde{z}$, and correlation length exponent, $\tilde{\nu}$, remain unchanged across the duality,
\be
\tilde\nu=\nu=1\,,\qquad\tilde{z}=z=1+\frac{16}{3\pi^2}\approx 1.5\,.
\ee
It should be possible to compute these exponents in a large-$N$ expansion of the Abelian Higgs model as well, and it would be interesting to compare the two results.
However, we caution that for $N>1$ the theories are no longer dual, and one limit may be more similar to the $N=1$ behavior than the other. It may also be possible to obtain exponents numerically for the dirty Abelian Higgs model with $N=1$.

Should the Abelian Higgs model with a random mass flow to such a QCP, this QCP will be characterized by a universal conductivity, which would be related to the universal conductivity of the fixed point we developed in Section~\ref{sec:RG} via Eq.~\eqref{eq: conductivity dictionary bosons}. We leave the calculation of the DC response of the Wilson-Fisher bosons with a random mass, both using a large-$N$ approach and numerical techniques, for future work. 

\subsection{Random Scalar and Vector Potentials}

We now consider the effects of perturbing by random scalar and vector potentials, as in Eq.~\eqref{eq:random-scalar-vector-potentials}.
The conclusion reached in that section only necessitated the preservation of a $\mathrm{U}(1)$ symmetry so our results remain valid even if the continuation to $N=1$ is invalid. By the mapping of charge to flux in Eq. \eqref{eq:charge-flux1}, the vortices $\tilde\phi$ experience a random scalar potential as a randomly sourced flux of $a_i$, 
\be
\mathcal{V}(\mathbf{x})\,J_0(\mathbf{x},t)\longleftrightarrow\frac{1}{2\pi}\mathcal{V}(\mathbf{x})\,\varepsilon^{ij}\pd_ia_j(\mathbf{x},t)\,.
\ee
Integrating by parts, we see that the disorder takes the form of a random current, $\mathcal{J}_i(\mathbf{x})=\pd_i\mathcal{V}/2\pi$. 
As demonstrated in Section~\ref{sec:CurrentDisorder}, the $\mathcal{V}(\vx)$ disorder is always relevant since it involves the temporal component of a conserved current, the flux $j^t=\varepsilon^{ij}\pd_ia_j/2\pi$. 
The ultimate fate of the Abelian Higgs theory is inaccessible through the perturbative RG approach employed throughout this paper.
Nevertheless, since we expect the $\phi$ bosons form a (perhaps glassy) insulating state in the presence of a random scalar potential, the conductivty dictionary in Eq.~\eqref{eq: conductivity dictionary bosons} indicates that the $\tilde{\phi}$ vortices have DC resistivity $\rho_{xx}(T/\omega\rightarrow0)\rightarrow0$. The vortices therefore appear to form a superconducting state. It would be interesting to better characterize this state in future work, using the conductivity dictionary and making suitable assumptions regarding fate of the Wilson-Fisher theory with a random scalar potential.

In keeping with the exchange of flux and charge, a random vector potential in the Wilson-Fisher theory maps to a random magnetic field $\mathcal{B}(\mathbf{x})=\varepsilon^{ij}\pd_i\mathcal{A}_j(\mathbf{x})$, which manifests as a random charge density in the Abelian Higgs model, 
\be
\mathcal{A}^i(\mathbf{x})\,J_i(\mathbf{x},t)\longleftrightarrow\frac{1}{2\pi}\mathcal{B}(\mathbf{x})\,a_t(\mathbf{x},t)\,.
\ee
As discussed in Section \ref{sec:CurrentDisorder}, this type of disorder is exactly marginal, leading to a line of fixed points parameterized by the dynamical exponent $z$, which depends on the disorder variance $\Delta_{\mathcal{A}}$.
\end{appendix}

\nocite{apsrev41Control}
\bibliographystyle{apsrev4-1}
\bibliography{random_WF}

\end{document}